\documentclass[letterpaper]{emulateapj}

\usepackage{epsfig}
\usepackage{amsmath}
\usepackage{epsfig,rotating}
\usepackage{natbib}
\usepackage{lscape}
\usepackage{enumerate}
\usepackage{multirow}
\usepackage{array}
\def\sles{\lower2pt\hbox{$\buildrel {\scriptstyle <}
   \over {\scriptstyle\sim}$}}

\def\sgreat{\lower2pt\hbox{$\buildrel {\scriptstyle >}
   \over {\scriptstyle\sim}$}}
\newcommand{\cN}[1]{\mathcal{N}}

\def\gsim{\;\rlap{\lower 2.5pt
 \hbox{$\sim$}}\raise 1.5pt\hbox{$>$}\;}
\def\lsim{\;\rlap{\lower 2.5pt
   \hbox{$\sim$}}\raise 1.5pt\hbox{$<$}\;}

\slugcomment{Submitted to ApJ, October 30, 2009}

\begin{document}

%%% Begin front material
%\twocolumn[%%% Begin front material

%\title{
%Coupled Radius-Orbit Evolutionary Models for the Inflated Planets
%WASP-4\lowercase{b} and WASP-12\lowercase{b}
%}

\title{
Explorations into the Viability of Coupled Radius-Orbit Evolutionary Models for Inflated Planets
}

\author{Laurent Ibgui\altaffilmark{1}, David S. Spiegel\altaffilmark{1}, and Adam Burrows\altaffilmark{1} }

\affil{$^1$Department of Astrophysical Sciences, Peyton Hall, Princeton University, Princeton, NJ 08544}

\vspace{0.5\baselineskip}

\email{ibgui@astro.princeton.edu, dsp@astro.princeton.edu, burrows@astro.princeton.edu}

\begin{abstract}
The radii of some transiting extrasolar giant planets are larger than would be expected by the 
standard theory. We address this puzzle with the model of coupled radius-orbit tidal evolution
developed by \citet{Ibgui_and_Burrows_2009}. The planetary radius is evolved self-consistently with
orbital parameters, under the influence of tidal torques and tidal dissipation in the interior of 
the planet. A general feature of this model, which we have previously demonstrated in the generic case,
is that a possible transient inflation of the planetary radius can 
temporarily interrupt its standard monotonic shrinking and can lead to the inflated radii that we
observe. In particular, a bloated planet with even a circular orbit may still be inflated due to an
earlier episode of tidal heating.
We have modified our model to include an orbital period dependence of the tidal dissipation factor
in the star, $Q'_{\ast} \propto P^{\gamma}$, $-1 \leqslant \gamma \leqslant 1$.
With this model, we search, for a tidally heated planet, orbital and radius evolutionary tracks
that fall within the observational limits of the radius, the semimajor axis, and the eccentricity of
the planet in its current estimated age range.
We find that, for some inflated planets (WASP-6b and WASP-15b), there are such tracks;
for another (TrES-4), there are none; and for still others (WASP-4b and WASP-12b), there are such 
tracks, but our model might imply that we are observing the planets at a special time.
Finally, we stress that there is a two to three order-of-magnitude timescale uncertainty of the inspiraling phase
of the planet into its host star, arising from uncertainties in the tidal dissipation factor in
the star $Q'_{\ast}$.
\end{abstract}

\keywords{planetary systems --- planets and satellites: general --- planets and satellites: individual (WASP-4b, WASP-6b, WASP-12b, WASP-15b, TrES-4)}
%%% End front material

%Examples of math-mode: $L_\sun = L_\odot$\\
%$M_J$ or $M_{\rm J}$

\section{Introduction}
\label{sec:intro}

The more than 60 transiting extrasolar giant planets (EGPs) discovered so far\footnote{
  See J. Schneider's Extrasolar Planet Encyclopaedia at
  http://exoplanet.eu, the Geneva Search Programme at
  http://exoplanets.eu, and the Carnegie/California compilation at
  http://exoplanets.org.} offer a unique opportunity to test and improve the
models of the structure and evolution of these bodies. 
The mass and radius of such planets can be inferred from a combination of radial velocity and 
transit lightcurve measurements that break the planet mass-inclination angle degeneracy.
A large theoretical effort has been undertaken for more than
a decade now to model and understand the evolution and the radii of transiting planets
\citep{guillot_et_al1996, burrows_et_al2000,
  Bodenheimer_et_al_2001, burrows_et_al2003, Baraffe_et_al_2003,
  Bodenheimer_et_al_2003,Gu_et_al_2003, burrows_et_al2004,
  Fortney_and_Hubbard_2004, Baraffe_et_al_2004, Chabrier_et_al_2004,
  laughlin_et_al_2005_1, Baraffe_et_al_2005, Baraffe_et_al_2006,
  burrows_et_al2007, Fortney_et_al_2007, Marley_et_al_2007,
  chabrier+baraffe2007, Liu_et_al_2008, Baraffe_et_al_2008,
  Ibgui_and_Burrows_2009, Miller_et_al_2009, Leconte_et_al_2009,
  Ibgui_et_al_2009_1}.

The radius of a gas giant planet depends on many physical effects that are particular to a given
planet-star system, including the mass and age of the planet;
the stellar irradiation flux and spectrum;
the composition -- in particular, the heavy-element content -- 
of the atmosphere, the envelope, and the core; the atmospheric circulation
that couples the day and the night sides; and any processes that could 
generate an extra power source in the interior of the planet, such as tidal heating.
Moreover, the transit radius effect \citep{burrows_et_al2003,Baraffe_et_al_2003} has
to be considered in order to infer the transit radius from the planet's physical radius.
Therefore, a custom evolutionary calculation is the most appropriate way to determine
a theoretical transit radius, in order to compare it with the observed transit radius.

The objective of the present paper is to test the coupled radius-orbit tidal evolution model 
developed by \citet{Ibgui_and_Burrows_2009} on some recently discovered inflated planets. We 
present evolutionary tracks for WASP-4b \citep{Wilson_et_al_2008, Southworth_et_al_2009_2,
Gillon_et_al_2009_1, Winn_et_al_2009_1} and WASP-12b \citep{Hebb_et_al_2009}.
We have also tested the model on TrES-4 \citep{Mandushev_et_al_2007, Sozzetti_et_al_2008},
WASP-6b \citep{Gillon_et_al_2009_2}, and WASP-15b \citep{West_et_al_2009}.
Some other inflated planets have been discovered recently, such as 
HAT-P-13b \citep{Bakos_et_al_2009_2}, WASP-17b \citep{Anderson_et_al_2009},
and CoRoT-5b \citep{Rauer_et_al_2009}. We might apply our formalism to them in the future.

The idea of exploring tidal heating as an explanation for the inflated radii was originally
formulated by \citet{Bodenheimer_et_al_2001}. They suggested an excitation mechanism to sustain a 
nonzero eccentricity, for example a planetary companion 
\citep{Bodenheimer_et_al_2003,Mardling_2007}.
To date, two transiting EGPs are known to be accompanied by a companion,
HAT-P-13b \citep{Bakos_et_al_2009_2} and HAT-P-7b \citep{Pal_et_al_2008,Winn_et_al_2009_4}.
This reinforces the plausibility of such a scenario.

\citet{Batygin_et_al_2009_2} have coupled a three-body tidal orbital evolution model with a model 
of the interior structure of HAT-P-13b. They found a quasi-stationary solution and possibly
consistent core masses, radii, and tidal heating rates. Assuming, as did \citet{Liu_et_al_2008}, 
that the systems are in a quasi-stationary state, \citet{Ibgui_et_al_2009_1} provide, for each of
the systems that they have studied (WASP-6b, WASP-12b, WASP-15b, TrES-4, HAT-P-12b) and for each of
the associated atmospheric opacities (solar, $10\times$solar) a relation between the heavy-element
core mass $M_{\rm core}$ and the ratio $e^2/Q'_{p}$, where $e$ is the orbital eccentricity and 
$Q'_{p}$ the tidal dissipation factor in the planet. This constraint results from a degeneracy 
between the dissipation heating rate in the interior of the planet (which increases the radius) and 
the mass of a possible heavy-element central core (which shrinks the radius).

For close-in EGPs (orbital separation $\sles~0.1-0.15$~AU), tidal torques are strong enough such 
that they can result in planetary orbital evolution, and they produce
tidal heating (dissipation) inside the planet. Such tidal effects were first suggested for 
transiting EGPs by \citet{Jackson_et_al_2008_1,Jackson_et_al_2008_2, Jackson_et_al_2008_3}.
Jackson et al. included the tides raised on the star and the tides raised on the planet. They found tidal 
rates close to the levels that \citet{burrows_et_al2007} proposed to maintain the observed radii of
some transiting EGPs. \citet{Ibgui_and_Burrows_2009} described a model that couples the two consequences
of these tidal effects -- planetary radius evolution and orbit evolution. They tested their
model on HD~209458b and found an explanation for the radius of this planet. Note that they also 
showed that a supersolar metallicity of the planetary atmosphere, without invoking tides, can  
explain the radius. \citet{Miller_et_al_2009} applied a similar method to all the transiting EGPs,
albeit with simplified models for the atmospheres of the planets and parent stars, and a restricted
range of possibilities for the tidal dissipation factors $Q'$. We have chosen a more detailed 
approach by adopting customized atmospheric models and an extended range for $Q'$. Therefore, due to
the complexity of the atmospheric calculations, we have selected a couple of planets. We believe 
that both approaches are complementary. The one followed by \citet{Miller_et_al_2009} provides a 
global estimate of the possibilities for matching the observed radii of the transiting EGPs, while our
approach, more precise and therefore applied to fewer planetary systems, focuses on more specific
issues such as the influence of the atmospheric opacity of the planet, a more detailed model for the
tidal dissipation in the star, and a phenomenological study of all the qualitatively possible 
behaviors of the evolutionary curves \citep{Ibgui_and_Burrows_2009}. The application of our model
to a subset of inflated transiting EGPs is the subject of this paper.

The paper outlines, in Section~\ref{sec:model}, the main assumptions of our coupled 
radius-orbit tidal evolution model, with a summary of the basic phenomenological results obtained from
the previous study by \citet{Ibgui_and_Burrows_2009}. It also explains our upgraded modeling of the
tidal dissipation factor in the star. Section~\ref{sec:M_Q_a_effect} 
demonstrates some additional generic results, such as the effect of $M_{\rm core}$, $Q'_{p}$, and the
initial semimajor axis. For each of the following planets, TrES-4, WASP-4b, WASP-6b, WASP-12b, and 
WASP-15b, we search for evolutionary tracks that fall within the observational limits of the 
radius, the semimajor axis, and the eccentricity of the planet in its current estimated age range. 
Our results are described in Section~\ref{sec:applications}. We have not found any such track for
TrES-4. We have found solutions for WASP-6b and WASP-15b. The cases of the planets WASP-4b and 
WASP-12b, for which we have coupled models that fit, are more interesting. Therefore, we present evolutionary
curves for these planets. In fact, the solutions that we obtain for these two planets are valid only 
for very short age ranges in comparison with the estimated ages of the planets. This would imply
that we are observing both planets at a very special time in their evolution, which would be a priori
unlikely. In Section~\ref{sec:plunging_timescale}, we discuss the plunging timescale of a planet into
its host star. Its uncertainty can span two to three orders of magnitude.
We summarize our results in Section~\ref{sec:conclusion}.\\

\section{The Coupled Radius-Orbit Evolutionary Model and the Period Dependence of the Tidal Dissipation Factor $Q'$}
\label{sec:model}

\subsection{The Phenomenology of the Coupled Radius-Orbit Evolution}
\label{subsec:phenomenology}

The model that we employ assumes a two-body gravitational and tidal interaction that consistently 
couples the evolution of the radius with the orbit of the planet.  It includes the tides raised on 
the planet and the tides raised on the star, along with stellar irradiation and detailed model 
atmospheres. The planetary radius evolves as a result of the competing influences of tidal 
heating in its convective interior (due to the dissipation of orbital energy) and radiative cooling
from its surface. The most interesting phenomenological result that we have obtained 
is that, for strong enough tides, the planet's radius can undergo a transient phase of inflation 
that temporarily interrupts its monotonic shrinking and 
resets its evolutionary clock \citep{Ibgui_and_Burrows_2009}. Moreover, we have demonstrated that,
due to thermal inertia, an earlier episode of tidal heating can result in an inflated radius at the
current age of the planet, even though its current orbit has nearly circularized.

\subsection{Formalism and Assumptions: a Summary}
\label{subsec:formalism}
The formalism, assumptions, and computational techniques have been
extensively described in \citet{Ibgui_and_Burrows_2009}. We summarize
them here.

The planetary structure consists of a gaseous ($\rm H_{2},He$)
isentropic envelope (helium mass fraction $Y=0.25$) described by the
equation of state of \citet{Saumon_et_al_1995}. It may also contain
an inner heavy-element core. The basic effect of the core is to shrink
the planetary radius \citep{burrows_et_al2007, Liu_et_al_2008, Ibgui_et_al_2009_1}.
However, in the context of coupled radius-orbit evolution, its
effect is more subtle, as will be discussed in \S\ref{sec:M_Q_a_effect}.
We restrict ourselves to a solar atmospheric opacity of the planet. 
The effects due to a higher opacity, namely an enhanced and accelerated
transient phase of radius inflation, have been described in \citet{Ibgui_and_Burrows_2009}.

The giant planet radius evolution is modeled with a Henyey code
\citep{Burrows_et_al_1993,Burrows_et_al_1997}. The radiative cooling
from the surface of the planet is linked to boundary conditions
\citep{burrows_et_al2003}. The latter incorporate realistic
irradiated planetary atmospheres calculated by CoolTLUSTY, a variant
of TLUSTY \citep{Hubeny_1988,Hubeny_and_Lanz_1995}. The specific
behaviors of such irradiated atmospheres are an active field of
research.  For example, an extra absorber in the upper-atmosphere may
produce a temperature inversion \citep{hubeny_et_al2003,
Burrows_et_al_2008_1, Burrows_et_al_2008_2, Fortney_et_al_2008_1,
Knutson_et_al_2008, Spiegel_et_al_2009_1}. We do not incorporate
such an extra absorber in the models of this paper. High metallicity
EGPs, such as Neptune-mass planets \citep{Spiegel_et_al_2009_2}, have
their own unique characteristics.  We calculate a customized host star spectrum
by interpolation of the \citet{Kurucz_1994} models at the actual
effective temperature and gravity of the star. Such a customized
approach can only improve the reliability of the model.

The equations governing the tidal evolution of the orbital
eccentricity $e$, the semimajor axis $a$, and the tidal heating rate
are the ones adopted by \citet{Ibgui_and_Burrows_2009}. We rewrite
them here for the sake of completeness \citep{Goldreich+Soter_1966,
  Kaula_1968, Peale_and_Cassen_1978, Murray_et_Dermott_1999,
  Bodenheimer_et_al_2001, Bodenheimer_et_al_2003, Gu_et_al_2004,
  Mardling_2007, Jackson_et_al_2008_2, Jackson_et_al_2008_3,
  Jackson_et_al_2008_4, Ferraz-Mello_et_al_2008, Barnes_et_al_2009_1}:
\begin{alignat}{2}
\frac{1}{e} \frac{de}{dt} &= - \frac{1}{a^{13/2}}  \left[ \overbrace{ \phantom{2} K_{1p_{\ast}} \frac{R_{p}^{5}}{Q'_{p}} \phantom{~e^{2}} } {}+
                                             \overbrace{ \phantom{ \frac{8}{25} \left( 1+\frac{57}{4}~e^{2} \right) {}} K_{2p_{\ast}} \frac{R_{\ast}^{5}}{Q'_{\ast}} } \right], \label{eq:e} \\
\frac{1}{a} \frac{da}{dt} &= - \frac{1}{a^{13/2}}  \left[\underbrace{  2 K_{1p_{\ast}}    \frac{R_{p}^{5}}{Q'_{p}}~e^{2} }_{\rm{tides~on~planet}} {}+
                                                      \underbrace{ \frac{8}{25} \left( 1+\frac{57}{4}~e^{2} \right) K_{2p_{\ast}} \frac{R_{\ast}^{5}}{Q'_{\ast}} }_{\rm{tides~on~star}}  \right]\, , \label{eq:a}
\end{alignat}
and
\begin{equation}
\dot{E}_{\rm tide}         = \phantom{-}  \left( \frac{63}{4} G^{3/2}M_{\ast}^{5/2} \right)  \frac{R_{p}^{5}}{Q'_{p}}  \frac{e^{2}}{a^{15/2}}\, , \label{eq:Etides}
\end{equation}
where $K_{1p_{\ast}}$ and $K_{2p_{\ast}}$ are constants defined by
\begin{eqnarray}
K_{1p_{\ast}} &= & \frac{63}{4} G^{1/2} \frac{M_{\ast}^{3/2}}{M_{p}},  \label{eq:K1p}\\
K_{2p_{\ast}} &= & \frac{225}{16} G^{1/2} \frac{M_{p}}{M_{\ast}^{1/2}} \label{eq:K2p}\ .
\end{eqnarray}
In the preceeding equations, $G$ is the gravitational constant,
$M_{p},~M_{\ast},R_{p}$, and $R_{\ast}$ are the masses and radii of the planet
and star, and $Q'_{p}$ and $Q'_{\ast}$ are the tidal dissipation factors in the 
planet and in the star \citep{Goldreich_1963,Goldreich+Soter_1966}. 
The planet radius is time-dependent, $R_{p}\left( t \right)$, but the star's radius is
assumed to be constant. The principal assumptions underpinning these
equations (more details can be found in \citealt{Ibgui_and_Burrows_2009}) are as follows.
We consider that this two-body interaction starts a few Myr after star formation,
precluding any interaction with a protoplanetary disk \citep{Goldreich_and_Sari_2003} 
or with other planets \citep{Ford_et_al_2003, Juric_and_Tremaine_2008,
  Chatterjee_et_al_2008, Ford_and_Rasio_2008, Nagasawa_et_al_2008}. We
also do not consider the Kozai interaction
\citep{Wu_and_Murray_2003,Wu_2003,Wu_et_al_2007,Nagasawa_et_al_2008}.
We assume that, after a few Myrs, the planet is close
enough to its host star that tidal effects can be
significant, typically using an ``initial" semimajor axis $a_{i}$ up to 0.1~AU
and we consider an ``initial" eccentricity $e_{i}$ that can range from
0.0 to 0.8.
We neglect stellar and planetary obliquities. We assume that the
planet's spin is synchronized (tidally locked) with its orbital
period, and that the star's spin rate is small compared with the orbital
mean motion. These equations are developed to lowest-order in $e$ \citep{Goldreich+Soter_1966}.
They become less good approximations
at higher values of eccentricity. We present results that include high values of $e$ for comparison
with previous work (\citealt{Jackson_et_al_2008_2,Ibgui_and_Burrows_2009,Miller_et_al_2009}
use the same approximation).
Higher-order terms may be considered in the future \citep{Mardling_and_Lin_2002}.
However, these equations rely on specific assumptions for the response of a body
to tidal forcing. Tidal theory is still an active research field with many
remaining ambiguities. Therefore, our approach is sufficient to 
describe the basic phenomenology.
These equations show that $e$ and $a$ can, given the above assumption,
only decrease.

Once the properties of both the star and the planet are fixed, the model
is controlled by four free parameters ($Q'_{p},Q'_{\ast},e_{i},a_{i}$)
that we vary in order to fit simultaneously, within the error bars of
the age of the planet, the planetary radius, the eccentricity, and the
semimajor axis. An important point to bear in mind is that the radius
and orbit evolution are strongly nonlinear and depend sensitively 
on these four free parameters. $e_{i}$ is varied from 0.00 to
0.80. $a_{i}$ is varied from its current (measured) value to 0.10~AU,
the approximate distance beyond which tidal effects are negligible in our cases.
The tidal dissipation in the planet is ruled by $Q'_{p}$, whose value
is poorly constrained. Experimental estimates provide $10^{5}$ to
$10^{6}$ for Jupiter \citep{Goldreich+Soter_1966, Yoder_and_Peale_1981},
while theoretical arguments suggest around
$10^{5}$ to $10^{7}$ \citep{Ogilvie_and_Lin_2004}, and up to $10^{8}$ for
planets with cores \citep{Goodman_and_Lackner_2009}. Consequently, we
have considered the range $10^5-10^8$ for $Q'_{p}$. A particular model
is specified for $Q'_{\ast}$ as is explained in the next subsection.

\subsection{The Tidal Dissipation Factor in the Star}
\label{subsec:Qstar}

The tides raised on the star play a specific role in that they are the
major if not the only contributor to the planet's orbital evolution, as soon as the
orbital eccentricity $e$ is small enough or is zero.
Moreover, for a zero eccentricity, the evolution of the semimajor axis
is described analytically, as was first demonstrated by \citep{Goldreich_1963}, as can also be
directly derived from eq.~(\ref{eq:a}):
\begin{equation}
a = a_{0}\left[ 1 - \frac{117}{4} \frac{G^{1/2}}{a_{0}^{13/2}} \frac{M_{p}}{M_{\ast}^{1/2}} ~ \frac{R_{\ast}^5}{Q'_{\ast}} \left(t-t_{0} \right)  \right]^{2/13} ,
\label{eq:a_when_e_equal_0}
\end{equation}
where $a_{0}$ is the semimajor axis at any time $t_{0}$ after the
eccentricity has become null.

The subsequent evolution of the transiting EGPs from their observed current 
state is a puzzling issue.
The stability, but also the timescale, of their evolution are being investigated
\citep{Spiegel_et_al_2009_3, Levrard_et_al_2009, Hellier_et_al_2009, Hamilton_2009}.
It is well known that the tides raised on the star ultimately cause the planet to 
spiral into its host star and probably eventually to be tidally disrupted
\citep{Rasio_et_al_1996,
Levrard_et_al_2009, Jackson_et_al_2009_1, Jackson_et_al_2009_2,Ibgui_and_Burrows_2009, Miller_et_al_2009}.
The associated timescale is higly dependent on $Q'_{\ast}$.

As an upgrade to the model used for our previous generic study
\citep{Ibgui_and_Burrows_2009}, we consider that the tidal dissipation
factor in the star, $Q'_{\ast}$, can evolve with the orbital period of
the system as follows:
\begin{equation}
Q'_{\ast} = 10^{\beta} \times  \left( \frac{P}{P_{0}} \right) ^\gamma , \label{eq:Qstar}
\end{equation}
where $P$ is the orbital period of the system, $P_{0}$ is a
reference orbital period (i.e. for which $Q'_{\ast}=10^{\beta}$). The
exponent $\gamma$ is between $-1$ and $+1$.  The theoretical
motivation for this range of $\gamma$ comes from the modeling of tidal
dissipation in fluid bodies and the modeling of turbulent kinematic
viscosity \citep{Zahn_1966,Zahn_1989,Goldreich_and_Nicholson_1977}.
The range of $\gamma$ has been empirically confirmed by
\citet{Spiegel_et_al_2009_3} who show, based on a statistical study of
the observed transiting planets' properties, that only such a range can
lead to a stationary rate of plunging planets into their host stars.
The range for $\beta$ is $5.0$~-~$8.0$, resulting in $Q'_{\ast}$
within $10^5-10^8$ at $P=P_{0}$. We adopt the same range as for
$Q'_{p}$.  We choose $P_{0}=10$~days for the reference orbital
period. Indeed, observational data from
\citet{Meibom_and_Mathieu_2005} on solar-type binaries in the open
cluster M35 suggest a value of $Q'_{\ast}\approx10^{6}$ for a period of
10~days \citep{Ogilvie_and_Lin_2007}.

\section{Coupled Evolution of the Tidally Heated Radius and the Orbit: the Effects of $M_{\lowercase{core}}$, 
         $Q'_{\lowercase{p}}$, and $\lowercase{a}_{\lowercase{i}}$}
\label{sec:M_Q_a_effect}

Major generic features of the radius and orbital co-evolution of a
close-in giant planet are described by \citet{Ibgui_and_Burrows_2009}.
Here, we present additional results: we evaluate the effect of $M_{\rm core}$, $Q'_{p}$, $a_{i}$ on the evolution of the radius
and the orbit of a transiting EGP. Our statements can be demonstrated with equations
(\ref{eq:e},\ref{eq:a},\ref{eq:Etides}). They have also been numerically tested on the
``generic transiting system'' employed by \citet{Ibgui_and_Burrows_2009},
namely the HD~209458 system. Here, the tides raised on the star are
neglected ($Q'_{\ast} \rightarrow \infty$) for clarity's sake.
Everything else being equal, increasing the values of either $M_{\rm core}$, $Q'_{p}$, or
$a_{i}$, results in delaying the appearance of the radius inflation peak and decreasing its value.
It also results in faster circularization of the orbit, and the final semimajor axis $a_{f}$ is
reached faster. As for the latter, by virtue of conservation of the angular momentum
of the system, $a_{f}$ is the same whatever (for whatever $M_{\rm core}$ or $Q'_{p}$),
but a larger $a_{i}$ results in a larger $a_{f}$.

The demonstration of the evolution of the radius inflation peak and the orbit 
is based on the fact that a higher value of each of the three parameters 
($M_{\rm core}$, $Q'_{p}$, $a_{i}$) results in a lower initial
tidal heating rate (see eq.~\ref{eq:Etides}).  It is straightforward for
an enhanced $Q'_{p}$ or $a_{i}$. \citet{burrows_et_al2007} established
that the larger the heavy-element core, the smaller the EGP radius.
At the same time, the evolutions of $e$ and $a$ start with a lower
rate. The tidal heating being initially lower, the ensuing transient
expansion phase is manifest to a lesser degree and later.

The conclusion is that, when tidal heating is coupled with 
orbital evolution, the addition of a core does not necessarily result in
a smaller radius.  The result depends on the age of the
system. In essence, at earlier age the planet with the larger core has
the smaller radius, but it is the opposite at later age.\\

\section{Coupled Evolution of the Tidally Heated Radius and the Orbit: Applications}
\label{sec:applications}
The validity of the model can be tested against the available data of the observed inflated 
transiting EGPs. Our objective is to find evolutionary tracks that fall within the observational
limits of the radius, the semimajor axis, and the eccentricity of the planet in its current estimated
age range. If we do find such tracks, we say that ``we fit the planet''. 
The first application was presented in \citet{Ibgui_and_Burrows_2009} for HD~209458b.
We were able to simultaneously fit the radius, the eccentricity, and the semimajor axis of this 
planet with the set ($Q'_{p},Q'_{\ast},e_{i},a_{i}$)= ($10^{6.55},10^{7.0},0.77,0.085~\rm AU$),
where $Q'_{\ast}$ is constant, and for a solar opacity. Note that if HD~209458b has 3 to 
$10\times$solar opacity, we can explain its radius without invoking the tidal heating argument.
In this paper, we present the results of our investigation for other inflated planets, WASP-4b 
(\S\ref{subsec:WASP-4b}) and WASP-12b (\S\ref{subsec:WASP-12b}). We assume a solar opacity, and no
central heavy-element core. We adopt a $Q'_{\ast}$ that varies according to eq.~(\ref{eq:Qstar})
with $\gamma=-1$ in order to smooth the ultimate plunging of the planet
into its host star.  We have also tested our coupled model for TrES-4, WASP-6b, and WASP-15b.
Observational data are listed in Table \ref{tab:transit_planets_data} for the planets' properties 
and in Table \ref{tab:host_stars_data} for the host stars' characteristics.

For each planet, we have tested a large number of combinations of the parameters 
($Q'_{p},Q'_{\ast},e_{i},a_{i}$). Recalling the $Q'_{\ast}$ evolutionary law given by 
eq.~(\ref{eq:Qstar}) in \S\ref{subsec:Qstar}, the reference orbital period is $P_{0}=10$~days.
We took 7 values for the $\beta$ parameter that determines $Q'_{\ast}$ at $P=P_{0}$ ($\beta=$ 5.0 to 8.0
in intervals of 0.5), 31 values for $Q'_{p}$ ($\log_{10}(Q'_{p})=$ 5.0 to 8.0 in intervals of 0.1),
81 values for $e$ (0.00 to 0.80, in intervals of 0.01), and a certain number 
(39 for WASP-4b and WASP-12b) of values of $a$ (the observed value up to 0.10, in intervals of 0.002). This 
represents $\sim$680,000 evolutionary curves tested for each planet at a given opacity. Given
the high nonlinearity of the evolutionary equations, and the sensitive dependence on these 
parameters (see \S\ref{subsec:formalism}), this approach represents a fairly exhaustive
exploration of all the possible combinations in order to select the ones that fit the
observed parameters.

In \S\ref{subsec:WASP-6b-15b_Tres-4}, we describe results of our models in the cases of
WASP-6b, WASP-15b, and TrES-4.
Perhaps the most interesting cases, however, are WASP-4b (\S\ref{subsec:WASP-4b} and
Fig.~\ref{fig:fig1}) and WASP-12b (\S\ref{subsec:WASP-12b} and
Fig.~\ref{fig:fig2}), which we can fit, assuming a solar opacity.

\subsection{WASP-6b, WASP-15b, TrES-4}
\label{subsec:WASP-6b-15b_Tres-4}

WASP-6b and WASP-15b were discovered by the Wide Angle Search for Planets (WASP) survey
\citep{Gillon_et_al_2009_2, West_et_al_2009}.  The radius of WASP-6b is 
$1.224^{+0.051}_{-0.052} R_{J}$, its mass is
$0.503^{+0.019}_{-0.038} M_{J}$, and its age is $11^{+7}_{-7}$~Gyr.
The radius of WASP-15b is
$1.428^{+0.077}_{-0.077} R_{J}$, its mass is
$0.542^{+0.050}_{-0.050} M_{J}$, and its age is $3.9^{+2.8}_{-1.3}$~Gyr.
These two planets can
easily be fit with little tidal heating ($\log_{10}(Q'_{p}) \geqslant 7.5$) at
solar opacity and for large age ranges.  Moreover, the opacity effect is
sometimes sufficient to explain some radii: WASP-6b can be fit with a
$3\times$solar opacity, without tidal heating \citep{Ibgui_et_al_2009_1}.
Evolutionary tracks (not shown) that fit these planets' observed properties look very similar to those of
HD~209458b, presented in \citet{Ibgui_and_Burrows_2009}.

TrES-4 was discovered by \citet{Sozzetti_et_al_2008}.  Its radius is
$1.783^{+0.093}_{-0.086} R_{J}$, its mass is
$0.925^{+0.081}_{-0.082} M_{J}$, and its age is $2.9^{+1.5}_{-0.4}$~Gyr.
This planet is one of the most inflated transiting planets known, and its size is difficult
to explain, even with tidal heating, at its estimated age.
We were not able to simultaneously fit the radius, eccentricity, and
semimajor axis of TrES-4, at any of the opacities that we have tested
(solar, $3\times$solar, $10\times$solar).

\subsection{WASP-4b}
\label{subsec:WASP-4b}

WASP-4b was discovered by \citet{Wilson_et_al_2008} and its parameters
were further refined by \citet{Southworth_et_al_2009_2},
\citet{Gillon_et_al_2009_1}, and \citet{Winn_et_al_2009_1}.  Its observed radius is
$1.365^{+0.021}_{-0.021} R_{J}$ and its mass is
$1.237^{+0.064}_{-0.064} M_{J}$ for an age of $6.5^{+2.3}_{-2.3}$~Gyr.
Its eccentricity is not well-constrained, with an upper limit of 0.096
\citep{Madhusudhan_and_Winn_2009}.

Examples of evolutionary curves that fit for WASP-4b are portrayed in Fig.~\ref{fig:fig1}.
This figure consists of four panels that depict, versus the age (in Gyr) of the planet, the simultaneous evolution
of its radius $R_{p}(R_{J})$ (top left panel), its eccentricity $e$
(top right), its semimajor axis $a\rm(AU)$ (bottom left), and
$\log_{10}(Q'_{\ast})$ (solid curves, left y axis) and the orbital period $P\rm(days)$ (dashed curves, right y axis)
in the bottom right.
Solar opacity is assumed for the planetary atmosphere. The radius
evolution without tides and a constant orbit is represented by a black
curve in the top left panel. It shows that the standard model
prediction is quite far from the measurement. The difference is
roughly $0.16~R_{J}$, i.e. $12\%$.  Among the tested $Q'_{\ast}$ at $P=P_{0}$
($\log_{10}[Q'_{\ast}(P_0)]=$ 5.0 to 8.0 in steps of 0.5), we
find fitting solutions for $\log_{10}[Q'_{\ast}(P_0)] \leqslant 6.5$.  The
smaller the $Q'_{\ast}$, the steeper the plunging slope of the
semimajor axis. The results we show in Fig.~\ref{fig:fig1} are among
the ones that have the smoothest plunging orbits. Moreover, for all
the fitting curves, $\log_{10}(Q'_{p})$  is greater than or equal to 6.9, which is a fairly
high value. In the examples on Fig.~\ref{fig:fig1}, $Q'_{p} =
10^{8.0}$.  Also, the initial eccentricity is large,
$e_{i}=0.80$. Solutions with lower $e_{i}$ exist (the lowest is 0.45),
but they are for $Q'_{\ast}(P_0) = 10^{5.0}$ or $10^{5.5}$, which are the
fastest plunging configurations. Three fitting evolutionary curves are
plotted, for three different $a_{i}$: 0.050, 0.052, 0.054~AU. We
identify, when $a_{i}$ increases, the delay of the appearance, and the
lower value, of the radius inflation peak, as stated in
\S\ref{sec:M_Q_a_effect}.  The curves end with thick dots, where
the periastron of the orbit reaches the Roche limit represented by a
brown horizontal line in the bottom left panel. The ranges of ages for
which simultaneous fits are obtained within the plotted $1\sigma$
measurement limits for $R_{p}$, $e$, $a$, are represented by two
orange vertical segments.  These ranges are very narrow ($\lesssim
0.2$~Gyr) in comparison with the estimated age of the planet,
$6.5^{+2.3}_{-2.3}$~Gyr. It is awkward to suggest that we observe this
planet at a very special transitional period in its life.
However, if we relax the fitting criterion (on $R_p$, $e$, and $a$) from 1$\sigma$ to 2$\sigma$ or 3$\sigma$,
there would be a wider range of ages for which the evolutionary tracks would fit the observed values.
Therefore, our models would no longer imply that the present is such a
special time in the life of this planet.
The bottom right panel shows the
evolution of $Q'_{\ast}$, which increases with time, from
$\log_{10}(Q'_{\ast}) \sim 6.8$ to $\sim 7.6$ over the whole
evolutionary path.  This is because of its inverse dependence on the
orbital period $P$, which decreases while the orbit circularizes,
from $\sim$$4.5$~days to $\sim$$0.8$~days at the Roche limit. The
fitting ranges are naturally at the measured period $P_{\rm measured}
\approx 1.3$~days for which $Q'_{\ast} \approx 10^{7.4}$. This
increase of $Q'_{\ast}$ corresponds to a decrease in the tidal
dissipation in the star and to a smoother evolution of its orbit. 
In the bottom left panel, we notice an apparent jump of the
slope in the semimajor axis evolution, previously pointed out and
justified by \citet{Miller_et_al_2009} in their simulaton of the
evolution of HD~209458b.  This jump is respectively at 5, 6, 8~Gyr for
$a_{i}=0.050, 0.052, 0.054$~AU. It can be explained quasi-analytically
using eq.~(\ref{eq:a}) in a more compact form:
\begin{equation}
\dot{a} = {\dot{a}}_{p}e^{2}+{\dot{a}}_{\ast}\left(1+\frac{57}{4}e^{2}\right),  \label{eq:adot}
\end{equation}
where $\dot a_{p}$ and $\dot a_{\ast}$ are the rates of evolution of
$a$ due to the tides, respectively raised on the planet and on the
star\footnote{Explicitly, the rates are defined by $\dot a_p=-2 K_{1p_{\ast}}\frac{R_p^{5}}{a^{11/2}Q'_p}$, and 
              $\dot a_{\ast}=-\frac{8}{25} {K_{2p_{\ast}}} \frac{R_\ast^{5}}{a^{11/2}Q'_\ast}$.}.
Before this apparent jump, both tides contribute to the
decrease of $a$, roughly by a comparable amount according to
numerical tests.  Then, during the rapid decrease of $e$ as shown in
top right panel of Fig.~\ref{fig:fig1}, the components of the sum
(\ref{eq:adot}) that depend on $e$ fall even more rapidly because of
the square dependence on $e$.  Finally, after this short transitional
phase, $a$ evolves only as ${\dot{a}}_{\ast}$, which is independent of
$e$, leading to the analytical evolution described by
eq.~(\ref{eq:a_when_e_equal_0}).

Figure~\ref{fig:fig1} shows that, once the orbit has circularized, the radius still
remains far above the value reached with no tides, as the top left panel
demonstrates at the end of evolution depicted by the thick dots.  This
specificity, already mentionned by \citet{Ibgui_and_Burrows_2009}, is
due to thermal inertia. In sum, it is possible that an inflated planet with a circular orbit can be explained 
with tidal heating.

\subsection{WASP-12b}
\label{subsec:WASP-12b}

WASP-12b, discovered by \citet{Hebb_et_al_2009}, is unique in many
respects.  It is the second-largest transiting planet to date, with a
radius of $1.79 ^{+0.09} _{-0.09} R_{J}$.  Its mass is $1.41 ^{+0.10
}_{-0.10 } M_{J}$. Its estimated eccentricity is
$e=0.049^{+0.015}_{-0.015}$. It is the most heavily irradiated
transiting EGP with a flux at the substellar point of 9.098$\rm \times
10^{9}~ergs~cm^{-2}~s^{-1}$.  It has one of the shortest orbital
periods, $P=1.09142 ~\rm days$.  Its orbit is very close to its Roche
limit, 0.0221~AU, while its periastron, $p=a(1-e)$, may be below this
limit: $0.0207 \lesssim p \lesssim 0.0229$ (see
Table~\ref{tab:transit_planets_data}).  The planet is perhaps on the
verge of being tidally disrupted. It may also be losing mass due
to Roche lobe overflow \citep{Gu_et_al_2003, Li_et_al_2009}. This
phenomenon is not modeled here.

In order to fit the radius, \citet{Miller_et_al_2009} invoke a floor
on the eccentricity and an extremely rapid expansion of the radius when
the planet starts to plunge. Our model provides fitting evolutionary
curves without imposing a floor on $e$ and without the rapid expansion. Figure~\ref{fig:fig2}, 
similar to Fig.~\ref{fig:fig1}, shows examples of fitting curves for a solar
opacity atmosphere.  As is done for WASP-4b, the radius evolution with no tides
is drawn in black in the top left panel.  The difference between the
radii is $\sim$$0.50~R_{J}$, that is $28\%$, much larger than for WASP-4b.
Similar to WASP-4b, we obtain solutions for $\log_{10}[Q'_{\ast}(P_0)] \leqslant 6.5$, and 
high values of $Q'_{p}$ are required, $Q'_{p} \geqslant 10^{7.2}$.
We show in Fig.~\ref{fig:fig2} three examples of
fitting evolutionary curves with the smoothest plunging orbits
($Q'_{\ast}(P_0)=10^{6.5}$), for three different $a_{i}$: 0.053, 0.055,
0.057~AU, and an initial eccentricity of $e_{i}=0.73$. Lower $e_{i}$
is possible, down to 0.53, but for lower $Q'_{\ast}(P_0)$ and, therefore,
faster plunging orbits. The delay of the appearance of the radius
peak and its lower value when $a_{i}$ increases are discernable.
The bottom left panel shows the evolution of the semimajor axis and how
close the planet is to its Roche limit. As for WASP-4b, the thick dots
indicate the end of the evolution where the periastron
reaches the Roche limit.  Despite the $P^{-1}$ dependence of
$Q'_{\ast}$ (bottom right panel) and, therefore, the smoother plunging
of the planet when the orbit has circularized (compared with the constant-$Q'_{\ast}$ case),
the eccentricity and the semimajor axis are decreasing
extremely fast, especially after 1~Gyr.  Thus, the age ranges
when the radius of the planet, the eccentricity, and the semimajor axis
simultaneously fit the measurements, are even narrower than for
WASP-4b.  Depicted by vertical orange segments, they are of the order
of 50~Myr.  Observing the planet in such a short interval of its
life, questionable for WASP-4b, is even less likely for WASP-12b.
Nevertheless, as we described for WASP-4b in \S\ref{subsec:WASP-4b}, using a 2$\sigma$ or 3$\sigma$ criterion for fitting
the observed properties of the system would similarly alleviate this problem.
Finally, as in the case of WASP-4b, the radius when the orbit has circularized is clearly
above the radius obtained with no tides.

\section{The Plunging Timescale Uncertainty}
\label{sec:plunging_timescale}

An intriguing issue about the transiting extrasolar giant planets
is their subsequent evolution. Tidal evolutionary equations show that
their fate is tidal disruption as they inspiral into their
host star \citep{Rasio_et_al_1996, Levrard_et_al_2009,
  Jackson_et_al_2009_1, Jackson_et_al_2009_2, Ibgui_and_Burrows_2009,
  Miller_et_al_2009}. Once the orbit has circularized, the plunging
timescale $\tau$ is described by an analytic expression, directly
resulting from the evolution of $a$ (eq.~\ref{eq:a_when_e_equal_0}),
where $P$ is the orbital period:
\begin{alignat}{2}
\tau &= & \quad \left(\frac{4}{117\phantom{\pi}} Q'_{\ast} \right) \left(\frac{a_{0}^{13/2} M_{\ast}^{1/2}}{G^{1/2} M_{p} R_{\ast}^5} \right) \\
     &= & \quad P \left( \frac{2}{{117\pi}} Q'_{\ast} \right)  \left( \frac{M_{\ast}}{M_{p}} \right) \left( \frac{a_{0}}{R_{\ast}} \right)^{5}.
\label{eq:tau_fall}
\end{alignat}

We define $\tau$ as the remaining time for the planet to fall into its
host star from the present time, assuming that its current
eccentricity is zero.  We thus choose in this section $t_{0}$, defined
in \S\ref{subsec:Qstar}, to be the present time, and therefore $a_{0}$
to be the measured semimajor axis.

This formula shows the sensitive dependence of $\tau$ on the current
semimajor axis of the system $a_{0}$. The
latter is, however, quite precisely determined by observations (see Table \ref{tab:transit_planets_data}).
At the same time, eq.~(\ref{eq:tau_fall}) exhibits a linear dependence of $\tau$ on
$Q'_{\ast}$, which is a very poorly constrained parameter.

To illustrate our point, we plot on Fig.~\ref{fig:fig3} the
theoretical evolution, starting from the present time, of the
semimajor axes $a$(AU) of two transiting EGPs, HD~209458b and
WASP-12b.  The time (in Myr) is represented logarithmically.
Our aim is to emphasize the dependence of these evolution on
$Q'_{\ast}$ that follows the generic law described by eq.~(\ref{eq:Qstar}).
 The reference orbital period $P_{0}$ is the observed period, $P_{0}=P_{\rm
  measured}$.  We explore three possibilities for $\gamma$, assumed to
range from -1 to +1 (see \S\ref{eq:a_when_e_equal_0}): 1(dotted
curves), 0(solid), -1 (dashed).  We also examine three possibilities
for the $\beta$ factor : 5(red curves), 6(blue), 7(green).  This
enables us to roughly encompass the current observational and
theoretical estimates of $Q'_{\ast}$.  The measured eccentricities of
these systems are low enough to consider the tides raised on the
planet to be negligible and, therefore, to ignore $Q'_{p}$.  The
analytical formula for the evolution of $a$ (eq.~\ref{eq:a_when_e_equal_0})
is applicable and so is the above
definition of the plunging timescale (eq.~\ref{eq:tau_fall}).  We
numerically checked this point by comparison with the integration of
the full evolutionary equations of \S\ref{subsec:formalism}.  The
Roche limits of both systems are plotted; they mark the end of the
evolutionary curves represented by thick dots.

The linear dependence of $\tau$ on $Q'_{\ast}$ directly affects the
evolutionary curves. For HD~209458b, the order of magnitude of the
plunging timescale can be 0.5~Gyr ($\beta=5$), 5~Gyr ($\beta=6$),
or 50~Gyr ($\beta=7$).  The dependence on the $\gamma$ parameter is
weaker, since the ratio $P/P_{\rm measured}$ remains around 1. The
trend is that a positive $\gamma$ results in the decrease of
$Q'_{\ast}$ while the orbital period $P$ of the planet decreases as
it spirals into its host star.  This results in an increase of the
tidal torque exerted on the planet and, therefore, in accelerated
plunging in comparison to the case with $\gamma=0$
($Q'_{\ast}$~constant) and a fortiori to the case with $\gamma=1$
($Q'_{\ast} \propto P^{-1}$). Including all these uncertainties,
HD~209458b can plunge between 0.5 and 60~Gyr from now, which is a
2-order-of-magnitude range. Note that \citet{Levrard_et_al_2009}
provide the evolution curve of HD~209458b for one of the cases
considered here: $\beta=6$ and $\gamma=1$, (blue dotted curve). Their
result is consistent with ours. By the same token, we find that
WASP-12b can plunge between 0.1 and 100~Myr from now, which is a
3-order-of-magnitude range. The much shorter timescale, in comparison
with that for HD~209458b, is mainly due to the ratio of the semimajor axis
($a=0.0471$~AU for HD~209458b and $a=0.0229$~AU for WASP-12b, see
Table~\ref{tab:transit_planets_data}) combined with the power
dependence $\tau \propto a_0^{13/2}$. Note that for WASP-12b, the
relative influence of the $\gamma$ parameter compared with that for the $\beta$
parameter is bigger than for HD~209458b. The evolutionary curves can
overlap.

This figure demonstrates that it is difficult to predict the evolution
of the orbits of transiting EGPs, given the poor knowledge of the
tidal dissipation factors in the host stars.  However, the recently
discovered WASP-18b \citep{Hellier_et_al_2009}, which has an orbital
period of only 0.94~days, might experience a measurable tidal decay time.
\citet{Hellier_et_al_2009} state that if $Q'_{\ast}=10^{6}$, the
epoch of transit would shift by 28~s after 10~yr.
There are several other effects that, in principle, could also cause changes in the timing properties of transits.
\citet{miralda-escude2002} points out that torques due to both the quadrupole moment of the star and the gravitational
perturbations from a hypothetical companion planet could cause precession of the orbital plane and of the planet's periapse,
with associated effects on transit time and duration.  Furthermore, \citet{Rafikov_2009} suggests that both general
relativistic apsidal precession and proper motion of the exoplanetary system with respect to our solar system could cause
similar or greater changes in transit time and duration.

\section{Conclusions and Discussion}
\label{sec:conclusion}

We have presented in this paper some new general results of the
coupled radius-orbit evolutionary model described in
\citet{Ibgui_and_Burrows_2009}, and we have applied the model to the
inflated planets WASP-4b and WASP-12b. We assumed a two-body gravitational and tidal
interaction between the planet and its host star, coupling the planetary
radius and the orbit evolution. We included the tides raised on the planet and
the tides raised on the star. Stellar irradiation
and a detailed planetary atmosphere are included. The fundamental
result is the transient inflation of the planetry radius that
temporarily interrupts its monotonic standard shrinking. An important
point is that even though the current orbit of the planet has almost
circularized, the radius of the planet can still be inflated due to an
earlier episode of tidal heating. This is why we stress that an
inflated planet with an observed circular orbit can still have 
tidal heating as an explanation of its radius. Fixing the planet
and star properties, the model is controlled by four free
parameters, ($Q'_{p},Q'_{\ast},e_{i},a_{i}$), that are the tidal
dissipation factors in the planet and in the star, and the initial
eccentricity and semimajor axis at the beginning of this two-body
evolution. We stress the sensitive and nonlinear dependence of the
evolutionary curves on these parameters.

We have demonstrated that an increase of either the core mass $M_{\rm core}$,
or $Q'_{p}$,
or $a_{i}$ results in a lower value of the radius inflation peak and
in a delay of its appearance.  The final semimajor axis is the same,
whatever $M_{\rm core}$ or $Q'_{p}$, but is larger when $a_{i}$ is
larger.  At an earlier age, the planet with the larger core has the
smaller radius, but this is opposite at later ages.

We have enhanced our model by including an orbital period dependence
of the tidal dissipation in the star, $Q'_{\ast} \propto P^{\gamma}$,
$-1 \leqslant \gamma \leqslant 1$.  $Q'_{\ast}$ drives the inspiral
of the planet into its host star.

Applications of our model to recently detected transiting inflated
planets show that:
\begin{itemize}\itemsep-0.04in%cm
\vspace{-6pt}
\item WASP-6b and WASP-15b can be fit at solar opacity over Gyr age ranges.
\item We have not found an acceptable fit for TrES-4, at either solar, $3\times$solar,
      or $10\times$solar planet atmospheric opacity.
\item WASP-4b can be fit at solar opacity with, for example, the
  combination ($Q'_{p},e_{i},a_{i}$)=($10^{8.0},0.80,0.050$) and
  with $Q'_{\ast}=10^{6.5}\times(P/\rm 10days)^{-1}$.
\item WASP-12b can be fit at solar opacity with, for example, the
  combination ($Q'_{p},e_{i},a_{i}$)=($10^{8.0},0.73,0.055$) and
  with $Q'_{\ast}=10^{6.5}\times(P/\rm 10days)^{-1}$.
\end{itemize}
\vspace{-6pt}
For WASP-4b and WASP-12b, the ranges of ages that allow simultaneous fits of radius, semimajor
axis, and orbital eccentricity, are very narrow, seeming to suggest that, if the two-body coupled evolutionary
model described herein is in fact responsible for these planets' inflated radii, then we are observing them at a
special epoch in their evolution.  However, relaxing the fit-criterion from 1$\sigma$ to 2$\sigma$ or 3$\sigma$
would alleviate this apparent problem.

Our results (in particular, for TrES-4) suggest that a coupled radius-orbit tidal evolution
model might not on its own explain the radii of all the inflated
transiting giant planets.  An alternative scenario with stationary
heating has been proposed \citep{Ibgui_et_al_2009_1} and applied to
all the planets discussed in this paper. Though not providing direct solutions to
the inflated radii issue, this scenario constrains the ratio $e^{2}/Q'_{p}$ for a
given $M_{\rm core}$. Finally, a combination of these two models could be imagined
with a two-body interaction, followed by a quasi-steady low eccentricity 
phase due to perturbations by a second planet.

The last point we make in this paper is the uncertainty of the
plunging timescale during the spiraling of the planet into its host
star. This timescale is strongly dependent on the semimajor axis;
specifically, it depends on $a$ to the 6.5 power. It also has a linear dependence on
$Q'_{\ast}$, which is a parameter that is uncertain by several orders of magnitude.
We have shown that HD~209458b can plunge in between 0.5 and 60~Gyr from now, a
2-order-of-magnitude range, and that WASP-12b can plunge in between 0.1 and
100~Myr from now, a 3-order-of-magnitude range.

\citet{Ibgui_and_Burrows_2009} have suggested caveats to, and ways to improve, the model employed here.
We close by noting several additional points.
The orbital evolution equations depend on the theory of tidal dissipation inside gaseous planets 
and stars.
Improvements to this theory might result in different evolutionary tracks (of $R_p$, $e$, and $a$) from the ones
presented in this paper.
Furthermore, we have noted that it is not strictly appropriate to apply these equations to
model scenarios with large values of orbital eccentricity.  However, both for comparison with previous
work and because the proper tidal theory remains unknown,
our present approach is a valuable step in exploring the extent to which tidal dissipation might 
explain the radii of the inflated EGPs.
Further observations that might help to constrain this model and to discriminate between this
and the stationary-state model of \citet{Ibgui_et_al_2009_1} include both increasing the accuracy of 
orbital eccentricity measurements and searching for companions to the transiting EGPs.
These and other advances will help us progress toward a better understanding of
the puzzle of the inflated planets.

\acknowledgements

We thank Ivan Hubeny for useful help on issues concerning the
computing of the atmospheric models for the boundary conditions. We
thank Jeremy Goodman for his instructive insights into the physical
modeling of the tidal dissipation factors. We thank Roman Rafikov for useful 
discussions related to the sources of change of transit time and duration.
We also thank Jason Nordhaus for useful discussions. The authors are pleased to
acknowledge that part of the work reported in this paper was
substantially performed at the TIGRESS high performance computer
center at Princeton University, which is jointly supported by the
Princeton Institute for Computational Science (PICSciE) and Engineering and the
Princeton University Office of Information Technology.
This study was supported by NASA grant NNX07AG80G and under
JPL/Spitzer Agreements 1328092, 1348668, and 1312647.

\bibliography{biblio}

\begin{thebibliography}{98}
\expandafter\ifx\csname natexlab\endcsname\relax\def\natexlab#1{#1}\fi

\bibitem[{{Anderson} {et~al.}(2009){Anderson}, {Hellier}, {Gillon}, {Triaud},
  {Smalley}, {Hebb}, {Collier Cameron}, {Maxted}, {Queloz}, {West}, {Bentley},
  {Enoch}, {Horne}, {Lister}, {Mayor}, {Parley}, {Pepe}, {Pollacco},
  {S{\'e}gransan}, {Udry}, \& {Wilson}}]{Anderson_et_al_2009}
{Anderson}, D.~R., {Hellier}, C., {Gillon}, M., {Triaud}, A.~H.~M.~J.,
  {Smalley}, B., {Hebb}, L., {Collier Cameron}, A., {Maxted}, P.~F.~L.,
  {Queloz}, D., {West}, R.~G., {Bentley}, S.~J., {Enoch}, B., {Horne}, K.,
  {Lister}, T.~A., {Mayor}, M., {Parley}, N.~R., {Pepe}, F., {Pollacco}, D.,
  {S{\'e}gransan}, D., {Udry}, S., \& {Wilson}, D.~M. 2009, submitted to \apj,
  arXiv:0908.1553

\bibitem[{{Bakos} {et~al.}(2009){Bakos}, {Howard}, {Noyes}, {Hartman},
  {Torres}, {Kovacs}, {Fischer}, {Latham}, {Johnson}, {Marcy}, {Sasselov},
  {Stefanik}, {Sipocz}, {Kovacs}, {Esquerdo}, {Pal}, {Lazar}, \&
  {Papp}}]{Bakos_et_al_2009_2}
{Bakos}, G.~A., {Howard}, A.~W., {Noyes}, R.~W., {Hartman}, J., {Torres}, G.,
  {Kovacs}, G., {Fischer}, D.~A., {Latham}, D.~W., {Johnson}, J.~A., {Marcy},
  G.~W., {Sasselov}, D.~D., {Stefanik}, R.~P., {Sipocz}, B., {Kovacs}, G.,
  {Esquerdo}, G.~A., {Pal}, A., {Lazar}, J., \& {Papp}, I. 2009, accepted to
  \apj, arXiv:0907.3525

\bibitem[{{Baraffe} {et~al.}(2006){Baraffe}, {Alibert}, {Chabrier}, \&
  {Benz}}]{Baraffe_et_al_2006}
{Baraffe}, I., {Alibert}, Y., {Chabrier}, G., \& {Benz}, W. 2006, \aap, 450,
  1221

\bibitem[{{Baraffe} {et~al.}(2008){Baraffe}, {Chabrier}, \&
  {Barman}}]{Baraffe_et_al_2008}
{Baraffe}, I., {Chabrier}, G., \& {Barman}, T. 2008, \aap, 482, 315

\bibitem[{{Baraffe} {et~al.}(2003){Baraffe}, {Chabrier}, {Barman}, {Allard}, \&
  {Hauschildt}}]{Baraffe_et_al_2003}
{Baraffe}, I., {Chabrier}, G., {Barman}, T.~S., {Allard}, F., \& {Hauschildt},
  P.~H. 2003, \aap, 402, 701

\bibitem[{{Baraffe} {et~al.}(2005){Baraffe}, {Chabrier}, {Barman}, {Selsis},
  {Allard}, \& {Hauschildt}}]{Baraffe_et_al_2005}
{Baraffe}, I., {Chabrier}, G., {Barman}, T.~S., {Selsis}, F., {Allard}, F., \&
  {Hauschildt}, P.~H. 2005, \aap, 436, L47

\bibitem[{{Baraffe} {et~al.}(2004){Baraffe}, {Selsis}, {Chabrier}, {Barman},
  {Allard}, {Hauschildt}, \& {Lammer}}]{Baraffe_et_al_2004}
{Baraffe}, I., {Selsis}, F., {Chabrier}, G., {Barman}, T.~S., {Allard}, F.,
  {Hauschildt}, P.~H., \& {Lammer}, H. 2004, \aap, 419, L13

\bibitem[{{Barnes} {et~al.}(2009){Barnes}, {Jackson}, {Raymond}, {West}, \&
  {Greenberg}}]{Barnes_et_al_2009_1}
{Barnes}, R., {Jackson}, B., {Raymond}, S.~N., {West}, A.~A., \& {Greenberg},
  R. 2009, \apj, 695, 1006

\bibitem[{{Batygin} {et~al.}(2009){Batygin}, {Bodenheimer}, \&
  {Laughlin}}]{Batygin_et_al_2009_2}
{Batygin}, K., {Bodenheimer}, P., \& {Laughlin}, G. 2009, \apjl, 704, L49

\bibitem[{{Bodenheimer} {et~al.}(2003){Bodenheimer}, {Laughlin}, \&
  {Lin}}]{Bodenheimer_et_al_2003}
{Bodenheimer}, P., {Laughlin}, G., \& {Lin}, D.~N.~C. 2003, \apj, 592, 555

\bibitem[{{Bodenheimer} {et~al.}(2001){Bodenheimer}, {Lin}, \&
  {Mardling}}]{Bodenheimer_et_al_2001}
{Bodenheimer}, P., {Lin}, D.~N.~C., \& {Mardling}, R.~A. 2001, \apj, 548, 466

\bibitem[{{Burrows} {et~al.}(2008{\natexlab{a}}){Burrows}, {Budaj}, \&
  {Hubeny}}]{Burrows_et_al_2008_1}
{Burrows}, A., {Budaj}, J., \& {Hubeny}, I. 2008{\natexlab{a}}, \apj, 678, 1436

\bibitem[{{Burrows} {et~al.}(2000){Burrows}, {Guillot}, {Hubbard}, {Marley},
  {Saumon}, {Lunine}, \& {Sudarsky}}]{burrows_et_al2000}
{Burrows}, A., {Guillot}, T., {Hubbard}, W.~B., {Marley}, M.~S., {Saumon}, D.,
  {Lunine}, J.~I., \& {Sudarsky}, D. 2000, \apjl, 534, L97

\bibitem[{{Burrows} {et~al.}(1993){Burrows}, {Hubbard}, {Saumon}, \&
  {Lunine}}]{Burrows_et_al_1993}
{Burrows}, A., {Hubbard}, W.~B., {Saumon}, D., \& {Lunine}, J.~I. 1993, \apj,
  406, 158

\bibitem[{{Burrows} {et~al.}(2007){Burrows}, {Hubeny}, {Budaj}, \&
  {Hubbard}}]{burrows_et_al2007}
{Burrows}, A., {Hubeny}, I., {Budaj}, J., \& {Hubbard}, W.~B. 2007, \apj, 661,
  502

\bibitem[{{Burrows} {et~al.}(2004){Burrows}, {Hubeny}, {Hubbard}, {Sudarsky},
  \& {Fortney}}]{burrows_et_al2004}
{Burrows}, A., {Hubeny}, I., {Hubbard}, W.~B., {Sudarsky}, D., \& {Fortney},
  J.~J. 2004, \apjl, 610, L53

\bibitem[{{Burrows} {et~al.}(2008{\natexlab{b}}){Burrows}, {Ibgui}, \&
  {Hubeny}}]{Burrows_et_al_2008_2}
{Burrows}, A., {Ibgui}, L., \& {Hubeny}, I. 2008{\natexlab{b}}, \apj, 682, 1277

\bibitem[{{Burrows} {et~al.}(1997){Burrows}, {Marley}, {Hubbard}, {Lunine},
  {Guillot}, {Saumon}, {Freedman}, {Sudarsky}, \& {Sharp}}]{Burrows_et_al_1997}
{Burrows}, A., {Marley}, M., {Hubbard}, W.~B., {Lunine}, J.~I., {Guillot}, T.,
  {Saumon}, D., {Freedman}, R., {Sudarsky}, D., \& {Sharp}, C. 1997, \apj, 491,
  856

\bibitem[{{Burrows} {et~al.}(2003){Burrows}, {Sudarsky}, \&
  {Hubbard}}]{burrows_et_al2003}
{Burrows}, A., {Sudarsky}, D., \& {Hubbard}, W.~B. 2003, \apj, 594, 545

\bibitem[{{Chabrier} \& {Baraffe}(2007)}]{chabrier+baraffe2007}
{Chabrier}, G., \& {Baraffe}, I. 2007, \apjl, 661, L81

\bibitem[{{Chabrier} {et~al.}(2004){Chabrier}, {Barman}, {Baraffe}, {Allard},
  \& {Hauschildt}}]{Chabrier_et_al_2004}
{Chabrier}, G., {Barman}, T., {Baraffe}, I., {Allard}, F., \& {Hauschildt},
  P.~H. 2004, \apjl, 603, L53

\bibitem[{{Chatterjee} {et~al.}(2008){Chatterjee}, {Ford}, {Matsumura}, \&
  {Rasio}}]{Chatterjee_et_al_2008}
{Chatterjee}, S., {Ford}, E.~B., {Matsumura}, S., \& {Rasio}, F.~A. 2008, \apj,
  686, 580

\bibitem[{{Ferraz-Mello} {et~al.}(2008){Ferraz-Mello}, {Rodr{\'{\i}}guez}, \&
  {Hussmann}}]{Ferraz-Mello_et_al_2008}
{Ferraz-Mello}, S., {Rodr{\'{\i}}guez}, A., \& {Hussmann}, H. 2008, Celestial
  Mechanics and Dynamical Astronomy, 101, 171

\bibitem[{{Ford} \& {Rasio}(2008)}]{Ford_and_Rasio_2008}
{Ford}, E.~B., \& {Rasio}, F.~A. 2008, \apj, 686, 621

\bibitem[{{Ford} {et~al.}(2003){Ford}, {Rasio}, \& {Yu}}]{Ford_et_al_2003}
{Ford}, E.~B., {Rasio}, F.~A., \& {Yu}, K. 2003, in Astronomical Society of the
  Pacific Conference Series, Vol. 294, Scientific Frontiers in Research on
  Extrasolar Planets, ed. D.~{Deming} \& S.~{Seager}, 181--188

\bibitem[{{Fortney} \& {Hubbard}(2004)}]{Fortney_and_Hubbard_2004}
{Fortney}, J.~J., \& {Hubbard}, W.~B. 2004, \apj, 608, 1039

\bibitem[{{Fortney} {et~al.}(2008){Fortney}, {Lodders}, {Marley}, \&
  {Freedman}}]{Fortney_et_al_2008_1}
{Fortney}, J.~J., {Lodders}, K., {Marley}, M.~S., \& {Freedman}, R.~S. 2008,
  \apj, 678, 1419

\bibitem[{{Fortney} {et~al.}(2007){Fortney}, {Marley}, \&
  {Barnes}}]{Fortney_et_al_2007}
{Fortney}, J.~J., {Marley}, M.~S., \& {Barnes}, J.~W. 2007, \apj, 659, 1661

\bibitem[{{Gillon} {et~al.}(2009{\natexlab{a}}){Gillon}, {Anderson}, {Triaud},
  {Hellier}, {Maxted}, {Pollaco}, {Queloz}, {Smalley}, {West}, {Wilson},
  {Bentley}, {Collier Cameron}, {Enoch}, {Hebb}, {Horne}, {Irwin}, {Joshi},
  {Lister}, {Mayor}, {Pepe}, {Parley}, {Segransan}, {Udry}, \&
  {Wheatley}}]{Gillon_et_al_2009_2}
{Gillon}, M., {Anderson}, D.~R., {Triaud}, A.~H.~M.~J., {Hellier}, C.,
  {Maxted}, P.~F.~L., {Pollaco}, D., {Queloz}, D., {Smalley}, B., {West},
  R.~G., {Wilson}, D.~M., {Bentley}, S.~J., {Collier Cameron}, A., {Enoch}, B.,
  {Hebb}, L., {Horne}, K., {Irwin}, J., {Joshi}, Y.~C., {Lister}, T.~A.,
  {Mayor}, M., {Pepe}, F., {Parley}, N., {Segransan}, D., {Udry}, S., \&
  {Wheatley}, P.~J. 2009{\natexlab{a}}, \aap, 501, 785

\bibitem[{{Gillon} {et~al.}(2009{\natexlab{b}}){Gillon}, {Smalley}, {Hebb},
  {Anderson}, {Triaud}, {Hellier}, {Maxted}, {Queloz}, \&
  {Wilson}}]{Gillon_et_al_2009_1}
{Gillon}, M., {Smalley}, B., {Hebb}, L., {Anderson}, D.~R., {Triaud},
  A.~H.~M.~J., {Hellier}, C., {Maxted}, P.~F.~L., {Queloz}, D., \& {Wilson},
  D.~M. 2009{\natexlab{b}}, \aap, 496, 259

\bibitem[{{Goldreich} \& {Nicholson}(1977)}]{Goldreich_and_Nicholson_1977}
{Goldreich}, P., \& {Nicholson}, P.~D. 1977, Icarus, 30, 301

\bibitem[{{Goldreich} \& {Sari}(2003)}]{Goldreich_and_Sari_2003}
{Goldreich}, P., \& {Sari}, R. 2003, \apj, 585, 1024

\bibitem[{{Goldreich} \& {Soter}(1966)}]{Goldreich+Soter_1966}
{Goldreich}, P., \& {Soter}, S. 1966, Icarus, 5, 375

\bibitem[{{Goldreich}(1963)}]{Goldreich_1963}
{Goldreich}, R. 1963, \mnras, 126, 257

\bibitem[{{Goodman} \& {Lackner}(2009)}]{Goodman_and_Lackner_2009}
{Goodman}, J., \& {Lackner}, C. 2009, \apj, 696, 2054

\bibitem[{{Gu} {et~al.}(2004){Gu}, {Bodenheimer}, \& {Lin}}]{Gu_et_al_2004}
{Gu}, P.-G., {Bodenheimer}, P.~H., \& {Lin}, D.~N.~C. 2004, \apj, 608, 1076

\bibitem[{{Gu} {et~al.}(2003){Gu}, {Lin}, \& {Bodenheimer}}]{Gu_et_al_2003}
{Gu}, P.-G., {Lin}, D.~N.~C., \& {Bodenheimer}, P.~H. 2003, \apj, 588, 509

\bibitem[{{Guillot} {et~al.}(1996){Guillot}, {Burrows}, {Hubbard}, {Lunine}, \&
  {Saumon}}]{guillot_et_al1996}
{Guillot}, T., {Burrows}, A., {Hubbard}, W.~B., {Lunine}, J.~I., \& {Saumon},
  D. 1996, \apjl, 459, L35

\bibitem[{{Hamilton}(2009)}]{Hamilton_2009}
{Hamilton}, D.~P. 2009, \nat, 460, 1086

\bibitem[{{Hartman}(2009)}]{Hartman_2009}
{Hartman}, J.~D. 2009, private communication

\bibitem[{{Hebb} {et~al.}(2009){Hebb}, {Collier-Cameron}, {Loeillet},
  {Pollacco}, {H{\'e}brard}, {Street}, {Bouchy}, {Stempels}, {Moutou},
  {Simpson}, {Udry}, {Joshi}, {West}, {Skillen}, {Wilson}, {McDonald},
  {Gibson}, {Aigrain}, {Anderson}, {Benn}, {Christian}, {Enoch}, {Haswell},
  {Hellier}, {Horne}, {Irwin}, {Lister}, {Maxted}, {Mayor}, {Norton}, {Parley},
  {Pont}, {Queloz}, {Smalley}, \& {Wheatley}}]{Hebb_et_al_2009}
{Hebb}, L., {Collier-Cameron}, A., {Loeillet}, B., {Pollacco}, D.,
  {H{\'e}brard}, G., {Street}, R.~A., {Bouchy}, F., {Stempels}, H.~C.,
  {Moutou}, C., {Simpson}, E., {Udry}, S., {Joshi}, Y.~C., {West}, R.~G.,
  {Skillen}, I., {Wilson}, D.~M., {McDonald}, I., {Gibson}, N.~P., {Aigrain},
  S., {Anderson}, D.~R., {Benn}, C.~R., {Christian}, D.~J., {Enoch}, B.,
  {Haswell}, C.~A., {Hellier}, C., {Horne}, K., {Irwin}, J., {Lister}, T.~A.,
  {Maxted}, P., {Mayor}, M., {Norton}, A.~J., {Parley}, N., {Pont}, F.,
  {Queloz}, D., {Smalley}, B., \& {Wheatley}, P.~J. 2009, \apj, 693, 1920

\bibitem[{{Hellier} {et~al.}(2009){Hellier}, {Anderson}, {Cameron}, {Gillon},
  {Hebb}, {Maxted}, {Queloz}, {Smalley}, {Triaud}, {West}, {Wilson}, {Bentley},
  {Enoch}, {Horne}, {Irwin}, {Lister}, {Mayor}, {Parley}, {Pepe}, {Pollacco},
  {Segransan}, {Udry}, \& {Wheatley}}]{Hellier_et_al_2009}
{Hellier}, C., {Anderson}, D.~R., {Cameron}, A.~C., {Gillon}, M., {Hebb}, L.,
  {Maxted}, P.~F.~L., {Queloz}, D., {Smalley}, B., {Triaud}, A.~H.~M.~J.,
  {West}, R.~G., {Wilson}, D.~M., {Bentley}, S.~J., {Enoch}, B., {Horne}, K.,
  {Irwin}, J., {Lister}, T.~A., {Mayor}, M., {Parley}, N., {Pepe}, F.,
  {Pollacco}, D.~L., {Segransan}, D., {Udry}, S., \& {Wheatley}, P.~J. 2009,
  \nat, 460, 1098

\bibitem[{{Hubeny}(1988)}]{Hubeny_1988}
{Hubeny}, I. 1988, Computer Physics Communications, 52, 103

\bibitem[{{Hubeny} {et~al.}(2003){Hubeny}, {Burrows}, \&
  {Sudarsky}}]{hubeny_et_al2003}
{Hubeny}, I., {Burrows}, A., \& {Sudarsky}, D. 2003, \apj, 594, 1011

\bibitem[{{Hubeny} \& {Lanz}(1995)}]{Hubeny_and_Lanz_1995}
{Hubeny}, I., \& {Lanz}, T. 1995, \apj, 439, 875

\bibitem[{{Ibgui} \& {Burrows}(2009)}]{Ibgui_and_Burrows_2009}
{Ibgui}, L., \& {Burrows}, A. 2009, \apj, 700, 1921

\bibitem[{{Ibgui} {et~al.}(2009){Ibgui}, {Burrows}, \&
  {Spiegel}}]{Ibgui_et_al_2009_1}
{Ibgui}, L., {Burrows}, A., \& {Spiegel}, D.~S. 2009, submitted to \apj,
  arXiv:0910.4394

\bibitem[{{Jackson} {et~al.}(2008{\natexlab{a}}){Jackson}, {Barnes}, \&
  {Greenberg}}]{Jackson_et_al_2008_4}
{Jackson}, B., {Barnes}, R., \& {Greenberg}, R. 2008{\natexlab{a}}, \mnras,
  391, 237

\bibitem[{{Jackson} {et~al.}(2009{\natexlab{a}}){Jackson}, {Barnes}, \&
  {Greenberg}}]{Jackson_et_al_2009_2}
---. 2009{\natexlab{a}}, \apj, 698, 1357

\bibitem[{{Jackson} {et~al.}(2008{\natexlab{b}}){Jackson}, {Greenberg}, \&
  {Barnes}}]{Jackson_et_al_2008_1}
{Jackson}, B., {Greenberg}, R., \& {Barnes}, R. 2008{\natexlab{b}}, in IAU
  Symposium, Vol. 249, IAU Symposium, ed. Y.-S. {Sun}, S.~{Ferraz-Mello}, \&
  J.-L. {Zhou}, 187--196

\bibitem[{{Jackson} {et~al.}(2008{\natexlab{c}}){Jackson}, {Greenberg}, \&
  {Barnes}}]{Jackson_et_al_2008_2}
{Jackson}, B., {Greenberg}, R., \& {Barnes}, R. 2008{\natexlab{c}}, \apj, 678,
  1396

\bibitem[{{Jackson} {et~al.}(2008{\natexlab{d}}){Jackson}, {Greenberg}, \&
  {Barnes}}]{Jackson_et_al_2008_3}
---. 2008{\natexlab{d}}, \apj, 681, 1631

\bibitem[{{Jackson} {et~al.}(2009{\natexlab{b}}){Jackson}, {Greenberg}, \&
  {Barnes}}]{Jackson_et_al_2009_1}
{Jackson}, B., {Greenberg}, R., \& {Barnes}, R. 2009{\natexlab{b}}, in Bulletin
  of the American Astronomical Society, Vol.~41, Bulletin of the American
  Astronomical Society, 491

\bibitem[{{Juri{\'c}} \& {Tremaine}(2008)}]{Juric_and_Tremaine_2008}
{Juri{\'c}}, M., \& {Tremaine}, S. 2008, \apj, 686, 603

\bibitem[{{Kaula}(1968)}]{Kaula_1968}
{Kaula}, W.~M. 1968, {An introduction to planetary physics - The terrestrial
  planets} (Space Science Text Series, New York: Wiley, 1968)

\bibitem[{{Knutson} {et~al.}(2008){Knutson}, {Charbonneau}, {Allen}, {Burrows},
  \& {Megeath}}]{Knutson_et_al_2008}
{Knutson}, H.~A., {Charbonneau}, D., {Allen}, L.~E., {Burrows}, A., \&
  {Megeath}, S.~T. 2008, \apj, 673, 526

\bibitem[{{Knutson} {et~al.}(2009){Knutson}, {Charbonneau}, {Burrows},
  {O'Donovan}, \& {Mandushev}}]{Knutson_et_al_2009_1}
{Knutson}, H.~A., {Charbonneau}, D., {Burrows}, A., {O'Donovan}, F.~T., \&
  {Mandushev}, G. 2009, \apj, 691, 866

\bibitem[{{Knutson} {et~al.}(2007){Knutson}, {Charbonneau}, {Noyes}, {Brown},
  \& {Gilliland}}]{knutson_et_al2007a}
{Knutson}, H.~A., {Charbonneau}, D., {Noyes}, R.~W., {Brown}, T.~M., \&
  {Gilliland}, R.~L. 2007, \apj, 655, 564

\bibitem[{{Kurucz}(1994)}]{Kurucz_1994}
{Kurucz}, R. 1994, Solar abundance model atmospheres for 0,1,2,4,8 km/s.~Kurucz
  CD-ROM No.~19.~ Cambridge, Mass.: Smithsonian Astrophysical Observatory,
  1994., 19

\bibitem[{{Laughlin} {et~al.}(2005){Laughlin}, {Wolf}, {Vanmunster},
  {Bodenheimer}, {Fischer}, {Marcy}, {Butler}, \&
  {Vogt}}]{laughlin_et_al_2005_1}
{Laughlin}, G., {Wolf}, A., {Vanmunster}, T., {Bodenheimer}, P., {Fischer}, D.,
  {Marcy}, G., {Butler}, P., \& {Vogt}, S. 2005, \apj, 621, 1072

\bibitem[{{Leconte} {et~al.}(2009){Leconte}, {Baraffe}, {Chabrier}, {Barman},
  \& {Levrard}}]{Leconte_et_al_2009}
{Leconte}, J., {Baraffe}, I., {Chabrier}, G., {Barman}, T., \& {Levrard}, B.
  2009, accepted to \aap, arXiv:0907.2669

\bibitem[{{Levrard} {et~al.}(2009){Levrard}, {Winisdoerffer}, \&
  {Chabrier}}]{Levrard_et_al_2009}
{Levrard}, B., {Winisdoerffer}, C., \& {Chabrier}, G. 2009, \apjl, 692, L9

\bibitem[{{Li} {et~al.}(2009){Li}, {Miller}, {Lin}, \&
  {Fortney}}]{Li_et_al_2009}
{Li}, S., {Miller}, N., {Lin}, D., \& {Fortney}, J. 2009, \nat, submitted

\bibitem[{{Liu} {et~al.}(2008){Liu}, {Burrows}, \& {Ibgui}}]{Liu_et_al_2008}
{Liu}, X., {Burrows}, A., \& {Ibgui}, L. 2008, \apj, 687, 1191

\bibitem[{{Madhusudhan} \& {Winn}(2009)}]{Madhusudhan_and_Winn_2009}
{Madhusudhan}, N., \& {Winn}, J.~N. 2009, \apj, 693, 784

\bibitem[{{Mandushev} {et~al.}(2007){Mandushev}, {O'Donovan}, {Charbonneau},
  {Torres}, {Latham}, {Bakos}, {Dunham}, {Sozzetti}, {Fern{\'a}ndez},
  {Esquerdo}, {Everett}, {Brown}, {Rabus}, {Belmonte}, \&
  {Hillenbrand}}]{Mandushev_et_al_2007}
{Mandushev}, G., {O'Donovan}, F.~T., {Charbonneau}, D., {Torres}, G., {Latham},
  D.~W., {Bakos}, G.~{\'A}., {Dunham}, E.~W., {Sozzetti}, A., {Fern{\'a}ndez},
  J.~M., {Esquerdo}, G.~A., {Everett}, M.~E., {Brown}, T.~M., {Rabus}, M.,
  {Belmonte}, J.~A., \& {Hillenbrand}, L.~A. 2007, \apjl, 667, L195

\bibitem[{{Mardling}(2007)}]{Mardling_2007}
{Mardling}, R.~A. 2007, \mnras, 382, 1768

\bibitem[{{Mardling} \& {Lin}(2002)}]{Mardling_and_Lin_2002}
{Mardling}, R.~A., \& {Lin}, D.~N.~C. 2002, \apj, 573, 829

\bibitem[{{Marley} {et~al.}(2007){Marley}, {Fortney}, {Hubickyj},
  {Bodenheimer}, \& {Lissauer}}]{Marley_et_al_2007}
{Marley}, M.~S., {Fortney}, J.~J., {Hubickyj}, O., {Bodenheimer}, P., \&
  {Lissauer}, J.~J. 2007, \apj, 655, 541

\bibitem[{{Meibom} \& {Mathieu}(2005)}]{Meibom_and_Mathieu_2005}
{Meibom}, S., \& {Mathieu}, R.~D. 2005, \apj, 620, 970

\bibitem[{{Miller} {et~al.}(2009){Miller}, {Fortney}, \&
  {Jackson}}]{Miller_et_al_2009}
{Miller}, N., {Fortney}, J.~J., \& {Jackson}, B. 2009, \apj, 702, 1413

\bibitem[{{Miralda-Escud{\'e}}(2002)}]{miralda-escude2002}
{Miralda-Escud{\'e}}, J. 2002, \apj, 564, 1019

\bibitem[{{Murray} \& {Dermott}(1999)}]{Murray_et_Dermott_1999}
{Murray}, C.~D., \& {Dermott}, S.~F. 1999, {Solar System Dynamics}, ed.
  C.~U.~P. (MD99)

\bibitem[{{Nagasawa} {et~al.}(2008){Nagasawa}, {Ida}, \&
  {Bessho}}]{Nagasawa_et_al_2008}
{Nagasawa}, M., {Ida}, S., \& {Bessho}, T. 2008, \apj, 678, 498

\bibitem[{{Ogilvie} \& {Lin}(2004)}]{Ogilvie_and_Lin_2004}
{Ogilvie}, G.~I., \& {Lin}, D.~N.~C. 2004, \apj, 610, 477

\bibitem[{{Ogilvie} \& {Lin}(2007)}]{Ogilvie_and_Lin_2007}
---. 2007, \apj, 661, 1180

\bibitem[{{P{\'a}l} {et~al.}(2008){P{\'a}l}, {Bakos}, {Torres}, {Noyes},
  {Latham}, {Kov{\'a}cs}, {Marcy}, {Fischer}, {Butler}, {Sasselov}, {Sip{\H
  o}cz}, {Esquerdo}, {Kov{\'a}cs}, {Stefanik}, {L{\'a}z{\'a}r}, {Papp}, \&
  {S{\'a}ri}}]{Pal_et_al_2008}
{P{\'a}l}, A., {Bakos}, G.~{\'A}., {Torres}, G., {Noyes}, R.~W., {Latham},
  D.~W., {Kov{\'a}cs}, G., {Marcy}, G.~W., {Fischer}, D.~A., {Butler}, R.~P.,
  {Sasselov}, D.~D., {Sip{\H o}cz}, B., {Esquerdo}, G.~A., {Kov{\'a}cs}, G.,
  {Stefanik}, R., {L{\'a}z{\'a}r}, J., {Papp}, I., \& {S{\'a}ri}, P. 2008,
  \apj, 680, 1450

\bibitem[{{Peale} \& {Cassen}(1978)}]{Peale_and_Cassen_1978}
{Peale}, S.~J., \& {Cassen}, P. 1978, Icarus, 36, 245

\bibitem[{{Rafikov}(2009)}]{Rafikov_2009}
{Rafikov}, R.~R. 2009, \apj, 700, 965

\bibitem[{{Rasio} {et~al.}(1996){Rasio}, {Tout}, {Lubow}, \&
  {Livio}}]{Rasio_et_al_1996}
{Rasio}, F.~A., {Tout}, C.~A., {Lubow}, S.~H., \& {Livio}, M. 1996, \apj, 470,
  1187

\bibitem[{{Rauer} {et~al.}(2009){Rauer}, {Queloz}, {Csizmadia}, {Deleuil},
  {Alonso}, {Aigrain}, {Almenara}, {Auvergne}, {Baglin}, {Barge}, {Borde},
  {Bouchy}, {Bruntt}, {Cabrera}, {Carone}, {Carpano}, {De la Reza}, {Deeg},
  {Dvorak}, {Erikson}, {Fridlund}, {Gandolfi}, {Gillon}, {Guillot}, {Guenther},
  {Hatzes}, {Hebrard}, {Kabath}, {Jorda}, {Lammer}, {Leger}, {Llebaria},
  {Magain}, {Mazeh}, {Moutou}, {Ollivier}, {Paetzold}, {Pont}, {Rabus},
  {Renner}, {Rouan}, {Shporer}, {Samuel}, {Schneider}, {Triaud}, \&
  {Wuchterl}}]{Rauer_et_al_2009}
{Rauer}, H., {Queloz}, D., {Csizmadia}, S., {Deleuil}, M., {Alonso}, R.,
  {Aigrain}, S., {Almenara}, J.~M., {Auvergne}, M., {Baglin}, A., {Barge}, P.,
  {Borde}, P., {Bouchy}, F., {Bruntt}, H., {Cabrera}, J., {Carone}, L.,
  {Carpano}, S., {De la Reza}, R., {Deeg}, H.~J., {Dvorak}, R., {Erikson}, A.,
  {Fridlund}, M., {Gandolfi}, D., {Gillon}, M., {Guillot}, T., {Guenther}, E.,
  {Hatzes}, A., {Hebrard}, G., {Kabath}, P., {Jorda}, L., {Lammer}, H.,
  {Leger}, A., {Llebaria}, A., {Magain}, P., {Mazeh}, T., {Moutou}, C.,
  {Ollivier}, M., {Paetzold}, M., {Pont}, F., {Rabus}, M., {Renner}, S.,
  {Rouan}, D., {Shporer}, A., {Samuel}, B., {Schneider}, J., {Triaud},
  A.~H.~M.~J., \& {Wuchterl}, G. 2009, accepted to \aap, arXiv:0909.3397

\bibitem[{{Saumon} {et~al.}(1995){Saumon}, {Chabrier}, \& {van
  Horn}}]{Saumon_et_al_1995}
{Saumon}, D., {Chabrier}, G., \& {van Horn}, H.~M. 1995, \apjs, 99, 713

\bibitem[{{Southworth} {et~al.}(2009){Southworth}, {Hinse}, {Burgdorf},
  {Dominik}, {Hornstrup}, {J{\o}rgensen}, {Liebig}, {Ricci}, {Th{\"o}ne},
  {Anguita}, {Bozza}, {Novati}, {Harps{\o}e}, {Mancini}, {Masi}, {Mathiasen},
  {Rahvar}, {Scarpetta}, {Snodgrass}, {Surdej}, \&
  {Zub}}]{Southworth_et_al_2009_2}
{Southworth}, J., {Hinse}, T.~C., {Burgdorf}, M.~J., {Dominik}, M.,
  {Hornstrup}, A., {J{\o}rgensen}, U.~G., {Liebig}, C., {Ricci}, D.,
  {Th{\"o}ne}, C.~C., {Anguita}, T., {Bozza}, V., {Novati}, S.~C.,
  {Harps{\o}e}, K., {Mancini}, L., {Masi}, G., {Mathiasen}, M., {Rahvar}, S.,
  {Scarpetta}, G., {Snodgrass}, C., {Surdej}, J., \& {Zub}, M. 2009, \mnras,
  399, 287

\bibitem[{{Sozzetti} {et~al.}(2009){Sozzetti}, {Torres}, {Charbonneau}, {Winn},
  {Korzennik}, {Holman}, {Latham}, {Laird}, {Fernandez}, {O'Donovan},
  {Mandushev}, {Dunham}, {Everett}, {Esquerdo}, {Rabus}, {Belmonte}, {Deeg},
  {Brown}, {Hidas}, \& {Baliber}}]{Sozzetti_et_al_2008}
{Sozzetti}, A., {Torres}, G., {Charbonneau}, D., {Winn}, J.~N., {Korzennik},
  S.~G., {Holman}, M.~J., {Latham}, D.~W., {Laird}, J.~B., {Fernandez}, J.,
  {O'Donovan}, F.~T., {Mandushev}, G., {Dunham}, E., {Everett}, M.~E.,
  {Esquerdo}, G.~A., {Rabus}, M., {Belmonte}, J.~A., {Deeg}, H.~J., {Brown},
  T.~N., {Hidas}, M.~G., \& {Baliber}, N. 2009, \apj, 691, 1145

\bibitem[{{Spiegel} {et~al.}(2009{\natexlab{a}}){Spiegel}, {Burrows}, {Ibgui},
  {Hubeny}, \& {Milsom}}]{Spiegel_et_al_2009_2}
{Spiegel}, D.~S., {Burrows}, A., {Ibgui}, L., {Hubeny}, I., \& {Milsom}, J.~A.
  2009{\natexlab{a}}, submitted to \apj, arXiv:0909.2043

\bibitem[{{Spiegel} {et~al.}(2009{\natexlab{b}}){Spiegel}, {Goodman},
  {Burrows}, \& {Ibgui}}]{Spiegel_et_al_2009_3}
{Spiegel}, D.~S., {Goodman}, J., {Burrows}, A., \& {Ibgui}, L.
  2009{\natexlab{b}}, \apj, in preparation

\bibitem[{{Spiegel} {et~al.}(2009{\natexlab{c}}){Spiegel}, {Silverio}, \&
  {Burrows}}]{Spiegel_et_al_2009_1}
{Spiegel}, D.~S., {Silverio}, K., \& {Burrows}, A. 2009{\natexlab{c}}, \apj,
  699, 1487

\bibitem[{{Torres} {et~al.}(2008){Torres}, {Winn}, \&
  {Holman}}]{Torres_et_al_2008}
{Torres}, G., {Winn}, J.~N., \& {Holman}, M.~J. 2008, \apj, 677, 1324

\bibitem[{{West} {et~al.}(2009){West}, {Anderson}, {Gillon}, {Hebb}, {Hellier},
  {Maxted}, {Queloz}, {Smalley}, {Triaud}, {Wilson}, {Bentley}, {Collier
  Cameron}, {Enoch}, {Horne}, {Irwin}, {Lister}, {Mayor}, {Parley}, {Pepe},
  {Pollacco}, {Segransan}, {Spano}, {Udry}, \& {Wheatley}}]{West_et_al_2009}
{West}, R.~G., {Anderson}, D.~R., {Gillon}, M., {Hebb}, L., {Hellier}, C.,
  {Maxted}, P.~F.~L., {Queloz}, D., {Smalley}, B., {Triaud}, A.~H.~M.~J.,
  {Wilson}, D.~M., {Bentley}, S.~J., {Collier Cameron}, A., {Enoch}, B.,
  {Horne}, K., {Irwin}, J., {Lister}, T.~A., {Mayor}, M., {Parley}, N., {Pepe},
  F., {Pollacco}, D., {Segransan}, D., {Spano}, M., {Udry}, S., \& {Wheatley},
  P.~J. 2009, \aj, 137, 4834

\bibitem[{{Wilson} {et~al.}(2008){Wilson}, {Gillon}, {Hellier}, {Maxted},
  {Pepe}, {Queloz}, {Anderson}, {Collier Cameron}, {Smalley}, {Lister},
  {Bentley}, {Blecha}, {Christian}, {Enoch}, {Haswell}, {Hebb}, {Horne},
  {Irwin}, {Joshi}, {Kane}, {Marmier}, {Mayor}, {Parley}, {Pollacco}, {Pont},
  {Ryans}, {Segransan}, {Skillen}, {Street}, {Udry}, {West}, \&
  {Wheatley}}]{Wilson_et_al_2008}
{Wilson}, D.~M., {Gillon}, M., {Hellier}, C., {Maxted}, P.~F.~L., {Pepe}, F.,
  {Queloz}, D., {Anderson}, D.~R., {Collier Cameron}, A., {Smalley}, B.,
  {Lister}, T.~A., {Bentley}, S.~J., {Blecha}, A., {Christian}, D.~J., {Enoch},
  B., {Haswell}, C.~A., {Hebb}, L., {Horne}, K., {Irwin}, J., {Joshi}, Y.~C.,
  {Kane}, S.~R., {Marmier}, M., {Mayor}, M., {Parley}, N., {Pollacco}, D.,
  {Pont}, F., {Ryans}, R., {Segransan}, D., {Skillen}, I., {Street}, R.~A.,
  {Udry}, S., {West}, R.~G., \& {Wheatley}, P.~J. 2008, \apjl, 675, L113

\bibitem[{{Winn} {et~al.}(2009{\natexlab{a}}){Winn}, {Holman}, {Carter},
  {Torres}, {Osip}, \& {Beatty}}]{Winn_et_al_2009_1}
{Winn}, J.~N., {Holman}, M.~J., {Carter}, J.~A., {Torres}, G., {Osip}, D.~J.,
  \& {Beatty}, T. 2009{\natexlab{a}}, \aj, 137, 3826

\bibitem[{{Winn} {et~al.}(2009{\natexlab{b}}){Winn}, {Johnson}, {Albrecht},
  {Howard}, {Marcy}, {Crossfield}, \& {Holman}}]{Winn_et_al_2009_4}
{Winn}, J.~N., {Johnson}, J.~A., {Albrecht}, S., {Howard}, A.~W., {Marcy},
  G.~W., {Crossfield}, I.~J., \& {Holman}, M.~J. 2009{\natexlab{b}}, \apjl,
  703, L99

\bibitem[{{Wu}(2003)}]{Wu_2003}
{Wu}, Y. 2003, in Astronomical Society of the Pacific Conference Series, Vol.
  294, Scientific Frontiers in Research on Extrasolar Planets, ed. D.~{Deming}
  \& S.~{Seager}, 213--216

\bibitem[{{Wu} \& {Murray}(2003)}]{Wu_and_Murray_2003}
{Wu}, Y., \& {Murray}, N. 2003, \apj, 589, 605

\bibitem[{{Wu} {et~al.}(2007){Wu}, {Murray}, \& {Ramsahai}}]{Wu_et_al_2007}
{Wu}, Y., {Murray}, N.~W., \& {Ramsahai}, J.~M. 2007, \apj, 670, 820

\bibitem[{{Yoder} \& {Peale}(1981)}]{Yoder_and_Peale_1981}
{Yoder}, C.~F., \& {Peale}, S.~J. 1981, Icarus, 47, 1

\bibitem[{{Zahn}(1966)}]{Zahn_1966}
{Zahn}, J.~P. 1966, Annales d'Astrophysique, 29, 489

\bibitem[{{Zahn}(1989)}]{Zahn_1989}
{Zahn}, J.-P. 1989, \aap, 220, 112

\end{thebibliography}

% ------------------------------------------------------------------------------------------------------------------------------------------------
% ------------------------------------------------------------ tables ----------------------------------------------------------------------------
% ------------------------------------------------------------------------------------------------------------------------------------------------
% |--- table 1 ---|
\clearpage
\begin{table}[ht]
\small
\begin{center}
\caption{Planet and Orbital Data} \label{tab:transit_planets_data}
\vspace{0.4in}
%\tablewidth{17.0cm}
%\begin{tabular}{p{3cm}cccccc}
\begin{tabular}{lccccccccc}
\hline\hline
%  \tableline % ou \hline
%  \tableline
\rule {0pt} {10pt}
  Planet          &            $a$                   &         $e$                  &    $P$        &         \multicolumn{2} {c}{$M_{p}$}                     &         $R_{p}$            &  $F_{p}$$^a$      & Roche limit &     References \\
                  &            (AU)                  &                              &   (days)      &        ($\rm M_{J}$)        &    ($\rm M_{\earth}$)      &      ($\rm R_{J}$)         &                   &    (AU)     &                \\
  \tableline \\[0.0cm]
   HD~209458b     &  $0.04707^{+0.00046}_{-0.00047}$ & $< 0.028                   $ &   $3.52475$   &  $0.685^{+0.015}_{-0.014}$  &  $218 ^{+5}_{-4}$          &  $1.320^{+0.024}_{-0.025}$ &    $1.000$        & 0.019       &      1,2,3     \\[0.2cm]
   TrES-4         &  $0.05105^{+0.00079}_{-0.00167}$ & $< 0.01                $$^c$ &   $3.55395$   &  $0.925^{+0.081}_{-0.082}$  &  $294 ^{+26}_{-26}$        &  $1.783^{+0.093}_{-0.086}$ &    $2.371$        & 0.024       &        4       \\[0.2cm]
   WASP-4b        &  $0.02340^{+0.00060}_{-0.00060}$ & $< 0.096                   $ &   $1.33823$   &  $1.237^{+0.064}_{-0.064}$  &  $393 ^{+20}_{-20}$        &  $1.365^{+0.021}_{-0.021}$ &    $1.706$        & 0.015       &       5,3      \\[0.2cm]
   WASP-6b        &  $0.0421 ^{+0.0008 }_{-0.0013 }$ & $  0.054^{+0.018}_{-0.015} $ &   $3.36101$   &  $0.503^{+0.019}_{-0.038}$  &  $160 ^{+6}_{-12}$         &  $1.224^{+0.051}_{-0.052}$ &    $0.463$        & 0.018       &        6       \\[0.2cm]
   WASP-12b       &  $0.0229 ^{+0.0008 }_{-0.0008 }$ & $  0.049^{+0.015}_{-0.015} $ &   $1.09142$   &  $1.41 ^{+0.10 }_{-0.10 }$  &  $448 ^{+32}_{-32}$        &  $1.79 ^{+0.09} _{-0.09} $ &    $9.098$        & 0.021       &        7       \\[0.2cm]
   WASP-15b       &  $0.0499 ^{+0.0018 }_{-0.0018 }$ & $  0.052^{+0.029}_{-0.040} $ &   $3.75207$   &  $0.542^{+0.050}_{-0.050}$  &  $172 ^{+16}_{-16}$        &  $1.428^{+0.077}_{-0.077}$ &    $1.696$        & 0.022       &        8       \\[0.2cm]
  \tableline
\end{tabular}
\bigskip
\tablenotetext{}{$^{a}$ $F_{p}$ is the stellar flux at the planet's substellar point, in units of $\rm 10^{9}~ergs~cm^{-2}~s^{-1}$.}
\tablenotetext{}{$^{b}$ According to \citet{Hartman_2009}, the 68.3$\%$ confidence upper limit is 0.065 and the 95.4$\%$ confidence upper limit is 0.085.}
\tablenotetext{}{$^{c}$ For specificity, we adopt this value of $0.01$. \citet{Knutson_et_al_2009_1} derive a $3 \sigma$ upper limit $|e\cos(\omega)|$ of 0.0058, where $\omega$ is the argument of periastron.}
\tablerefs{
(1) \citet{knutson_et_al2007a},        %---HD209458b
(2) \citet{Torres_et_al_2008},         %---HD209458b
(3) \citet{Madhusudhan_and_Winn_2009}, %---HD209458b, WASP-4b, eccentricity
(4) \citet{Sozzetti_et_al_2008},       %---TrES-4
(5) \citet{Winn_et_al_2009_1},         %---WASP-4b
(6) \citet{Gillon_et_al_2009_2},       %---WASP-6b
(7) \citet{Hebb_et_al_2009},           %---WASP-12b
(8) \citet{West_et_al_2009}.           %---WASP-15b
}
\end{center}
\end{table}

\vspace{2cm}

% |--- table 2 ---|
\begin{table}[ht]
\small
\begin{center}
\caption{Host Star Data} \label{tab:host_stars_data}
\vspace{0.2in}

\begin{tabular}{p{3cm}ccccc}
\hline\hline
%  \tableline
%  \tableline
\rule {0pt} {10pt}
  Star               &          $M_{\ast}$           &            $R_{\ast}$           &        $T_{\ast}$         &        $\left[ Fe/H \right]_{\ast}$     &        Age          \\
                     &        $\rm M_{\sun}$         &          $\rm R_{\sun}$         &          (K)              &                (dex)                    &       (Gyr)         \\
  \tableline \\[0.0cm]
   HD~209458         &  $1.101^{+0.066}_{-0.062}$    &    $1.125^{+0.020}_{-0.023}$    &    $6065^{+50}_{-50}$     &         $0.00^{+0.05}_{-0.05}$          &    $3.1^{+0.8}_{-0.7}$   \\[0.2cm]
   TrES-4            &  $1.404^{+0.066}_{-0.134}$    &    $1.846^{+0.096}_{-0.087}$    &    $6200^{+75 }_{-75 }$   &         $+0.14^{+0.09}_{-0.09}$         &    $2.9^{+1.5}_{-0.4}$   \\[0.2cm]
   WASP-4            &  $0.925^{+0.040}_{-0.040}$    &    $0.912^{+0.013}_{-0.013}$    &    $5500^{+100}_{-100}$   &         $-0.03^{+0.09}_{-0.09}$         &    $6.5^{+2.3}_{-2.3}$   \\[0.2cm]
   WASP-6            &  $0.880^{+0.050}_{-0.080}$    &    $0.870^{+0.025}_{-0.036}$    &    $5450^{+100}_{-100}$   &         $-0.20^{+0.09}_{-0.09}$         &    $11 ^{+7  }_{-7  }$   \\[0.2cm]
   WASP-12           &  $1.35 ^{+0.14 }_{-0.14 }$    &    $1.57 ^{+0.07 }_{-0.07 }$    &    $6300^{+200}_{-100}$   &         $+0.30^{+0.05}_{-0.15}$         &    $2  ^{+1  }_{-1  }$   \\[0.2cm]
   WASP-15           &  $1.18 ^{+0.12 }_{-0.12 }$    &    $1.477^{+0.072}_{-0.072}$    &    $6300^{+100}_{-100}$   &         $-0.17^{+0.11}_{-0.11}$         &    $3.9^{+2.8}_{-1.3}$   \\[0.2cm]
  \tableline
\tablenotetext{}{Notes $-$ The references are the same as the ones used for the data in Table {\ref{tab:transit_planets_data}}. The ages are less well constrained and should be taken with caution.}

\end{tabular}
%\vspace{-0.4cm}
%\tablecomments{Data are from \citet{knutson_et_al2007a}, \citet{Torres_et_al_2008}, and \citet{Madhusudhan_and_Winn_2009}.}
%\tablenotetext{1}{These data are from \citet{Torres_et_al_2008}}
\end{center}
\end{table}
% ------------------------------------------------------------------------------------------------------------------------------------------------
% ---------------------------------------------------------- end tables --------------------------------------------------------------------------
% ------------------------------------------------------------------------------------------------------------------------------------------------

% ------------------------------------------------------------------------------------------------------------------------------------------------
% ------------------------------------------------------------ figures ---------------------------------------------------------------------------
% ------------------------------------------------------------------------------------------------------------------------------------------------

% |--- fig 1 ---|
\clearpage
\begin{landscape}
\begin{figure}[ht]
%\plotone{ms_evolution_fig1.eps}
%\centering\includegraphics[scale=0.80]{old_eps/ms_evolution_fig1.eps}
\centerline{
\includegraphics[width=12.0cm,angle=0,clip=true]{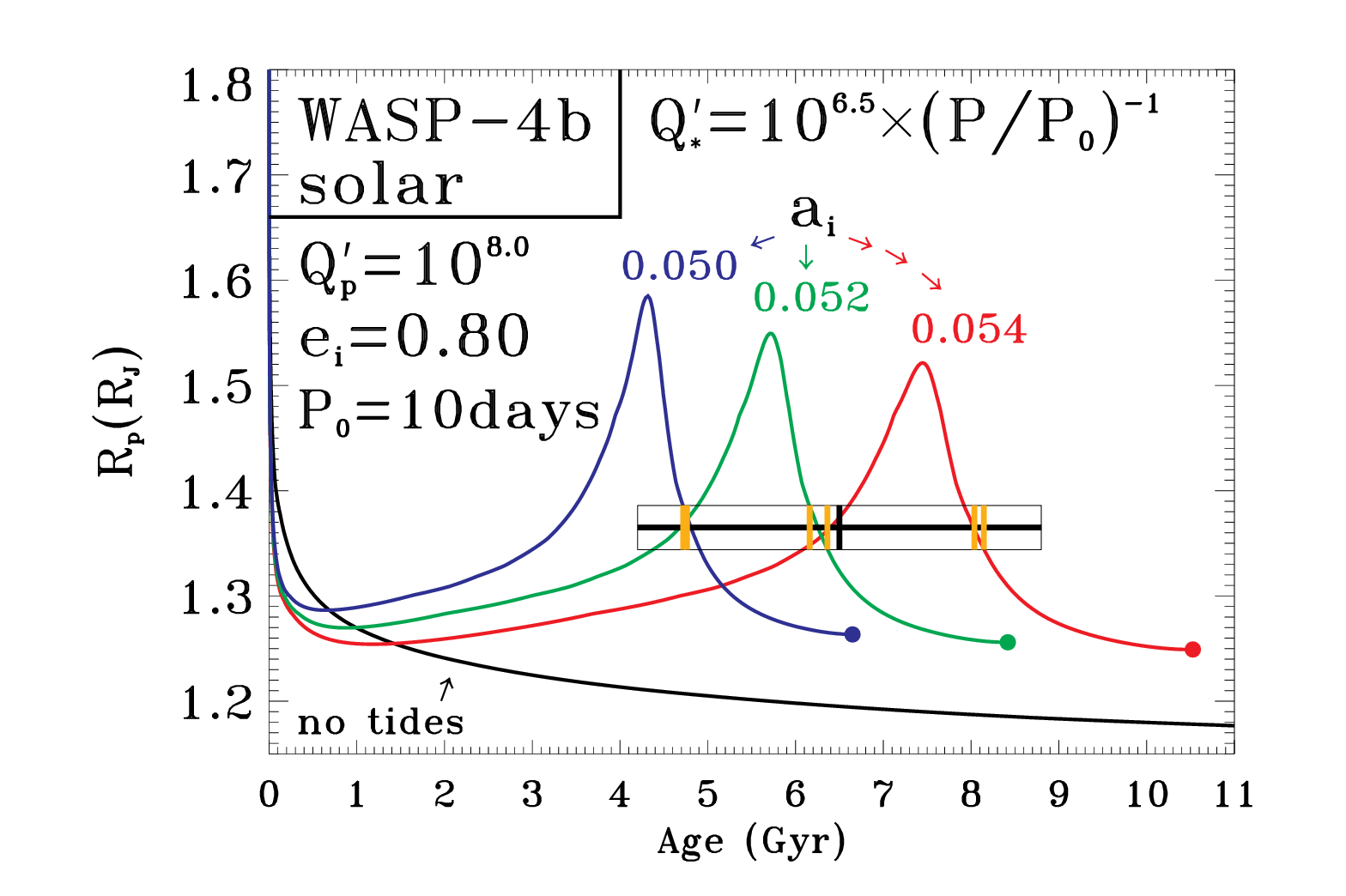}
\includegraphics[width=12.0cm,angle=0,clip=true]{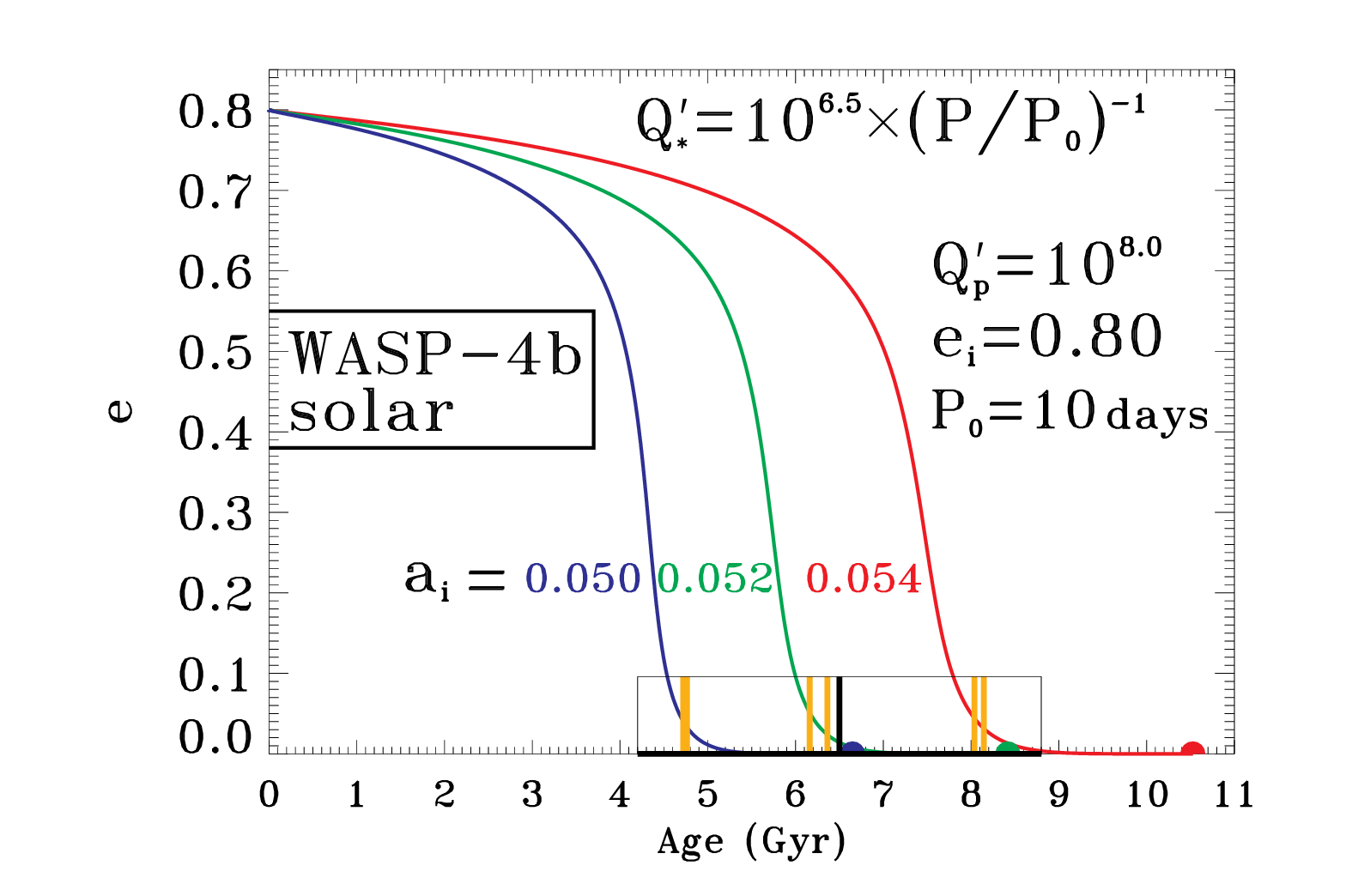}}  %width=17.0cm
\centerline{
\includegraphics[width=12.0cm,angle=0,clip=true]{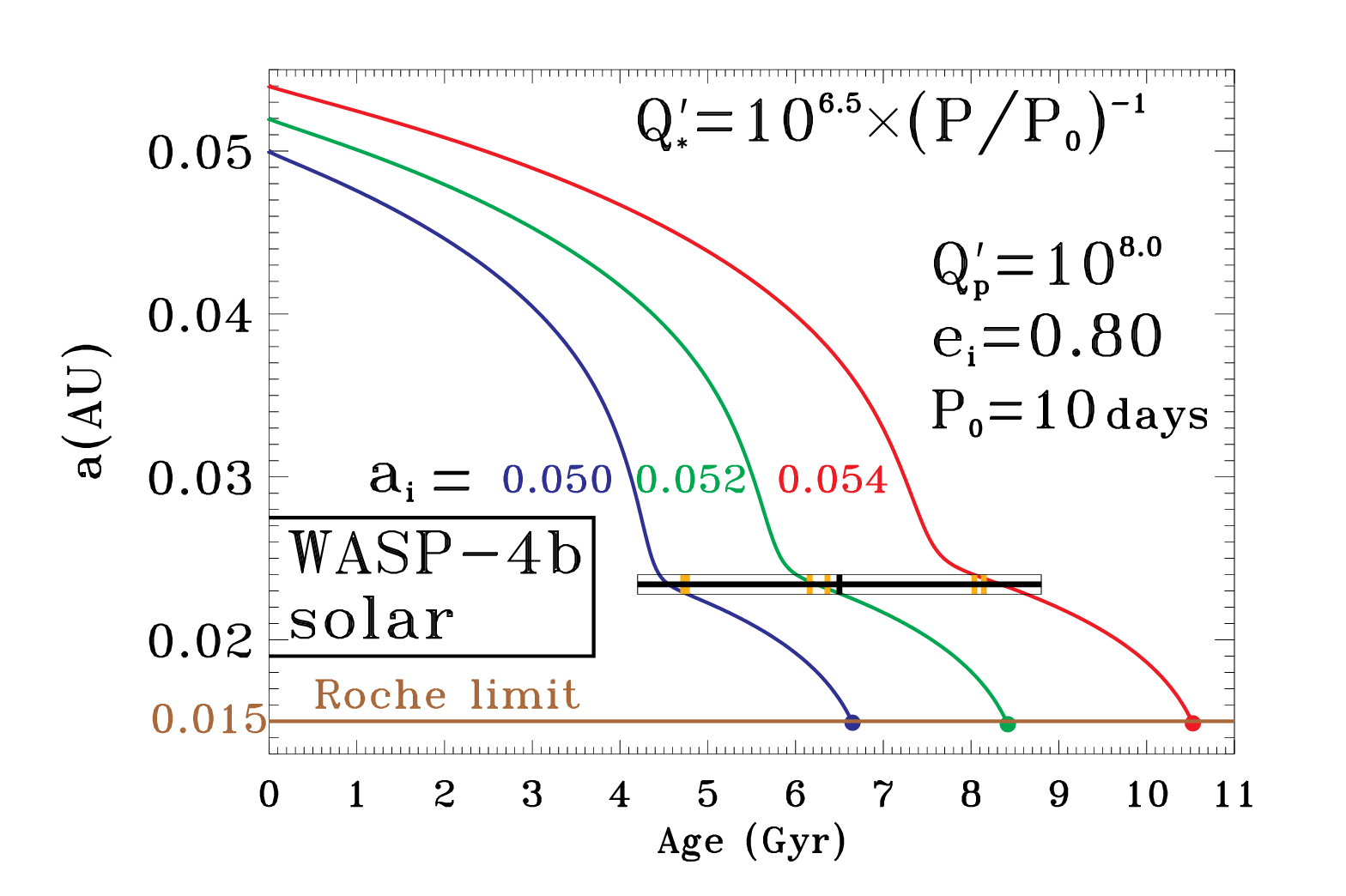}
\includegraphics[width=12.0cm,angle=0,clip=true]{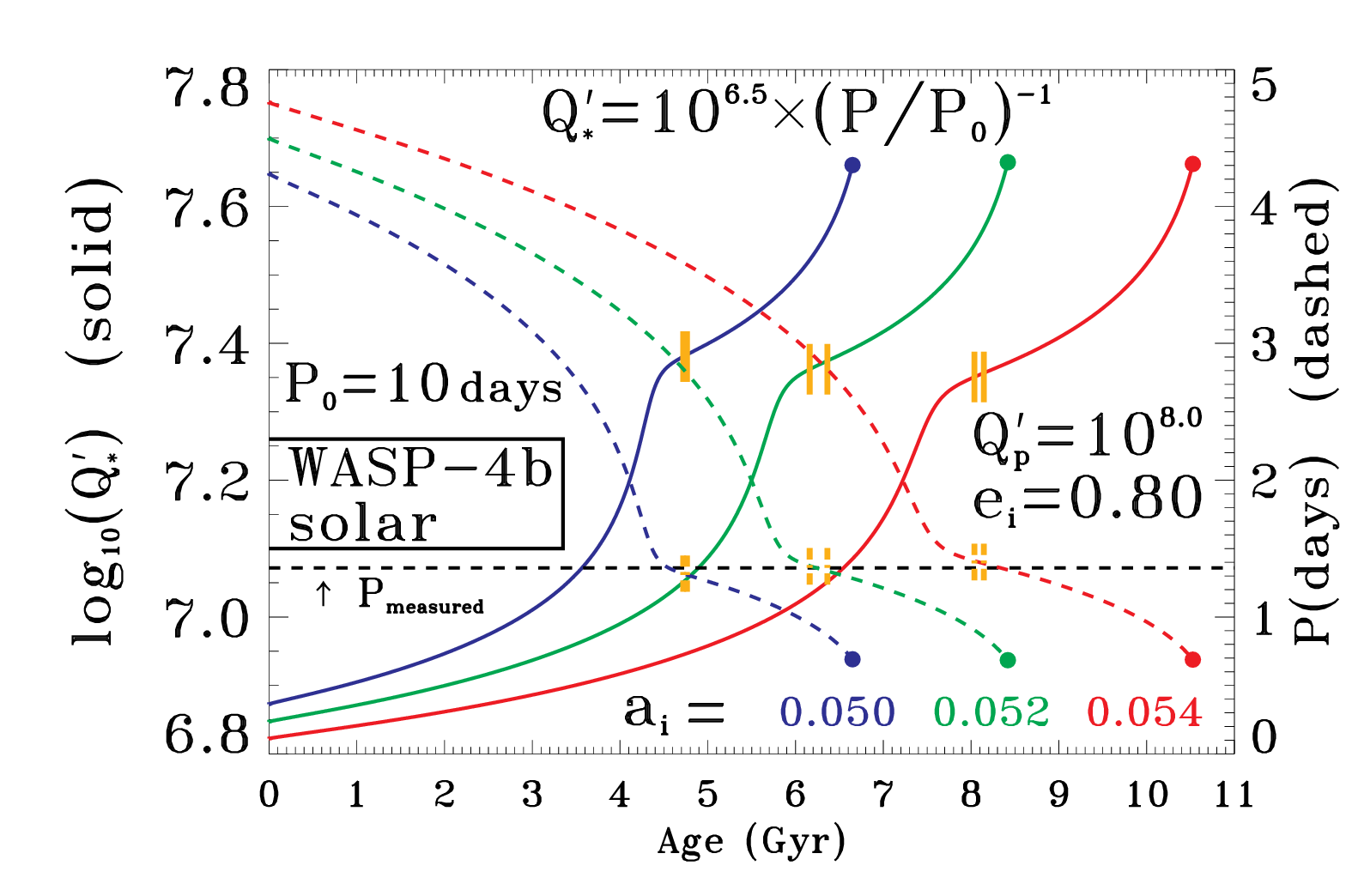}}
%%\rotatebox{90}{test}
\caption{\footnotesize Examples of fitting (within 1$\sigma$ of the measurements)
         evolutionary curves for WASP-4b at solar atmospheric opacity, assuming  
         coupled evolution with tides of the planetary radius and orbit \citep{Ibgui_and_Burrows_2009}.
         Top left, top right, and bottom left show (versus the age in Gyr) the co-evolution of the planetary radius $R_{\rm p}(R_{\rm J})$,
         the orbital eccentricity $e$, and the semimajor axis $a$(AU). Moreover, bottom right are plotted $\rm {log_{10}} (Q'_{\ast})$ (solid, units on left axis),
         where $Q'_{\ast}$ is the tidal dissipation factor for the star, and the orbital period $P$(days) (dashed, units on right axis).
         The pairs of orange vertical segments define the ranges of ages for which the fits are simultaneously obtained for $R_{p}$, $e$, and $a$,
         whose measured values are plotted along with the 1$\sigma$ limits.
         Notice how narrow these ranges are ($\lesssim$~0.2~Gyr) compared with the wide range of possibilities for the age of the system ($\sim$4.6~Gyr).
         Three cases are plotted. They differ in the initial semimajor axis $a_{i}$= 0.050(blue), 0.052(green), 0.054(red). For the three of them, the initial 
         eccentricity is $e_{i}$=0.80, the tidal dissipation factor for the planet is a constant $Q'_{p}=10^{8.0}$,
         and $Q'_{\ast}$ evolves as $P^{-1}$ with $Q'_{\ast}=10^{6.5}$ at $P=10$~days.
         Thick dots end the evolutionary curves at the age, for each case, when the periastron of the orbit $p=a(1-e)$ reaches the Roche limit, drawn in bottom left.
         Bottom right, the dashed black horizontal curve indicates the measured orbital period of the system. See Section~\ref{sec:applications} for a discussion.
        }
\label{fig:fig1}
\end{figure}

% |--- fig 2 ---|
\clearpage
%\begin{landscape}
\begin{figure}[ht]
%\plotone{ms_evolution_fig1.eps}
%\centering\includegraphics[scale=0.80]{old_eps/ms_evolution_fig1.eps}
\centerline{
\includegraphics[width=12.0cm,angle=0,clip=true]{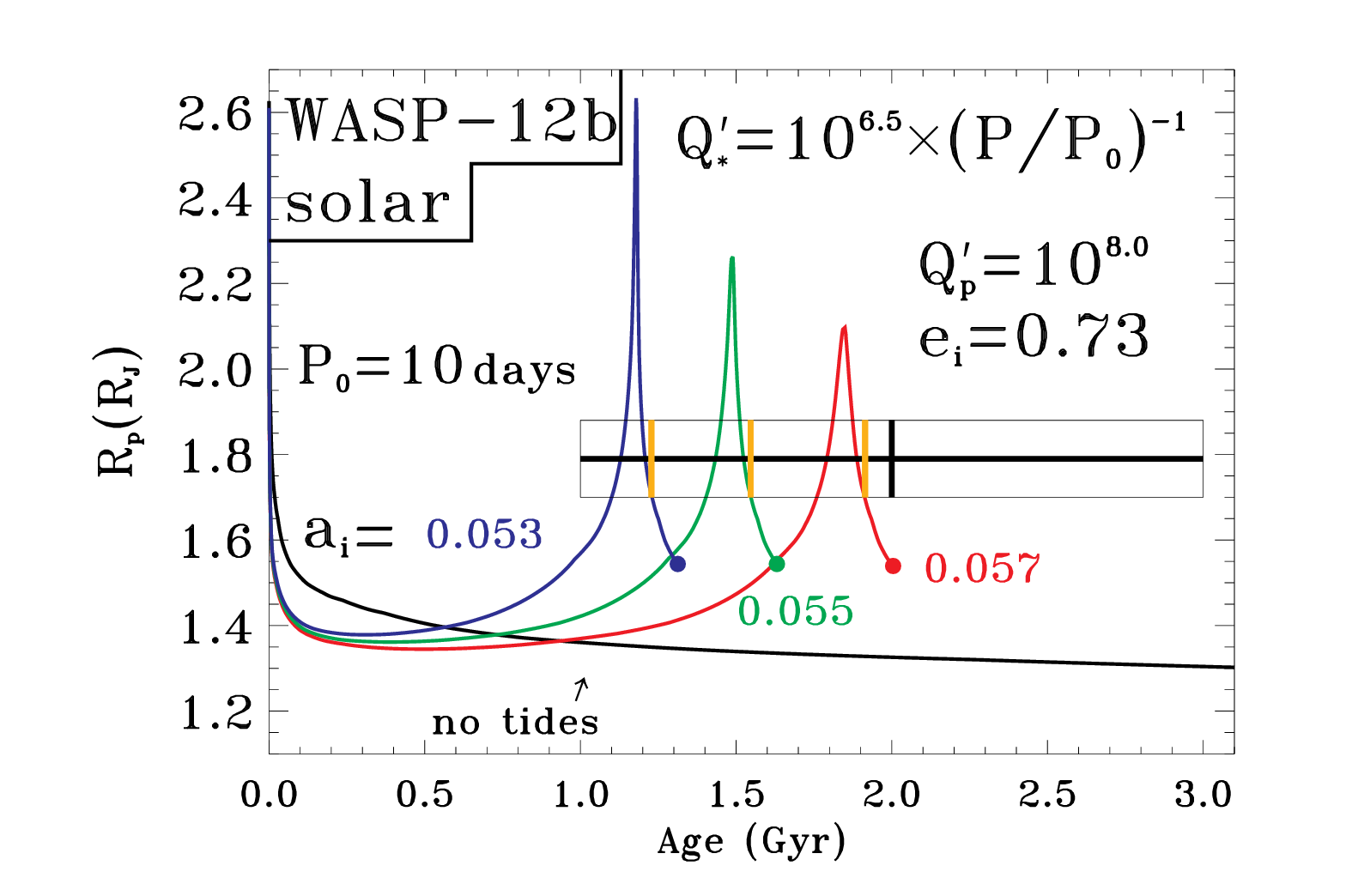}
\includegraphics[width=12.0cm,angle=0,clip=true]{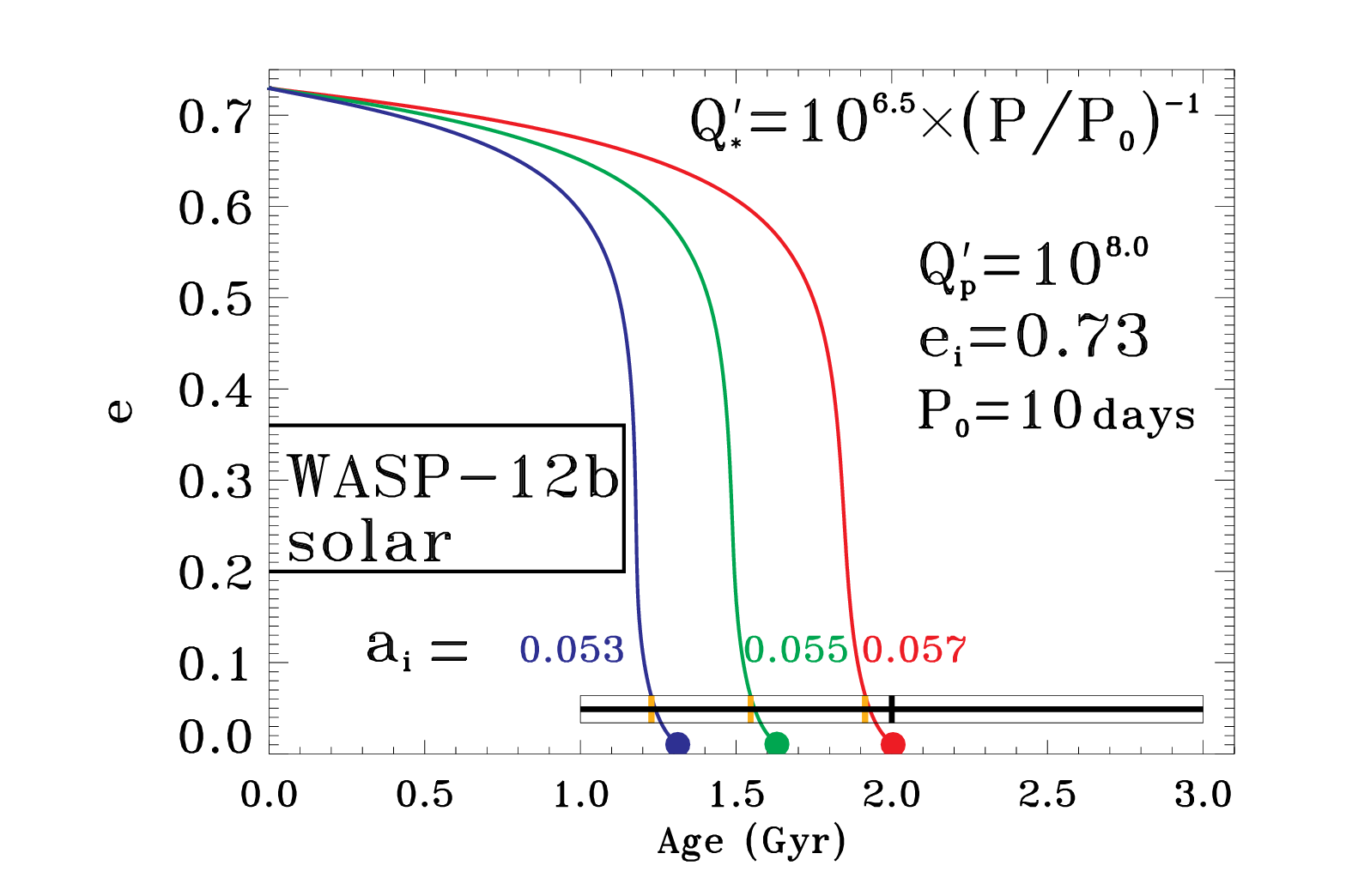}}  %width=17.0cm
\centerline{
\includegraphics[width=12.0cm,angle=0,clip=true]{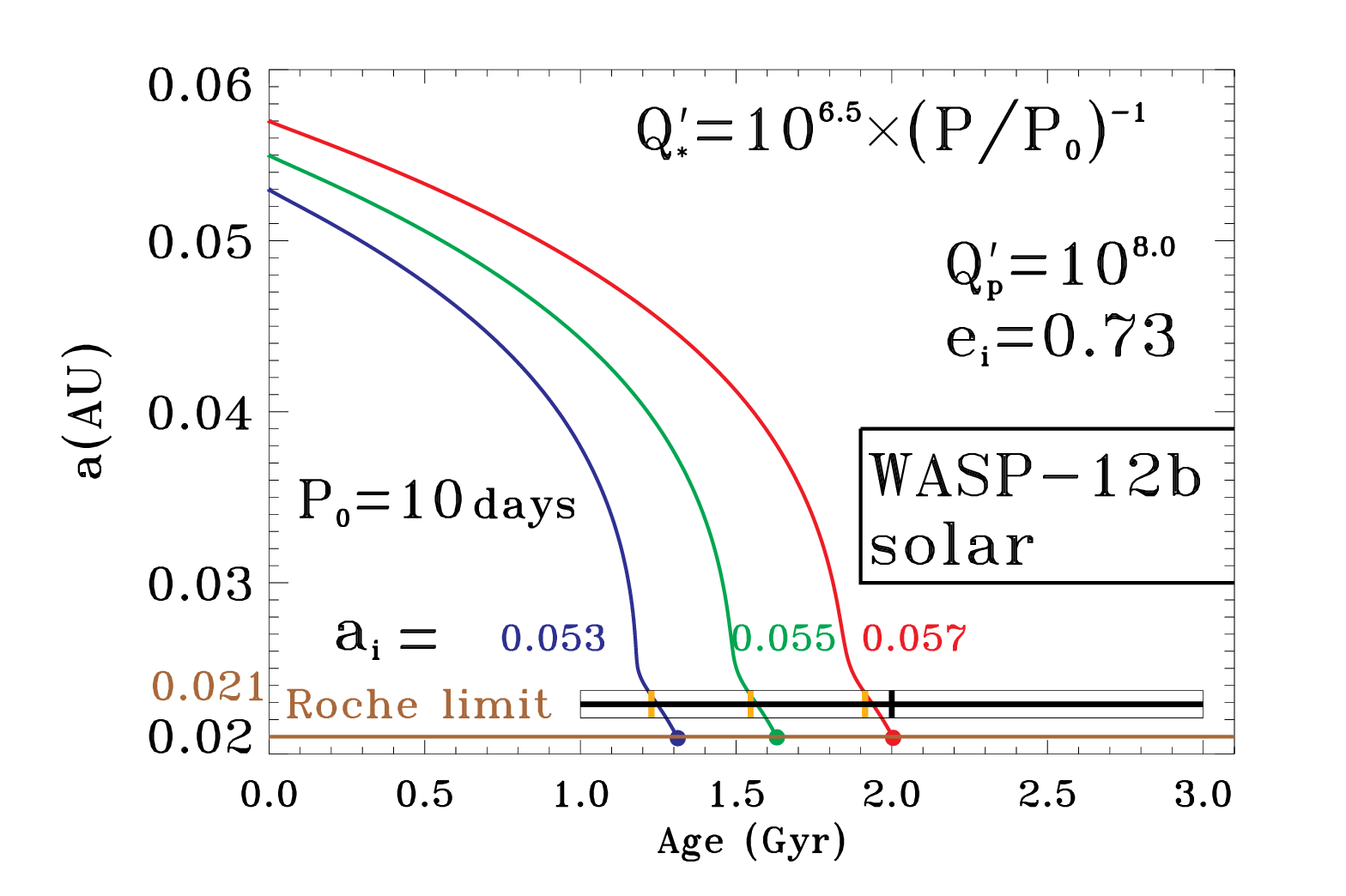}
\includegraphics[width=12.0cm,angle=0,clip=true]{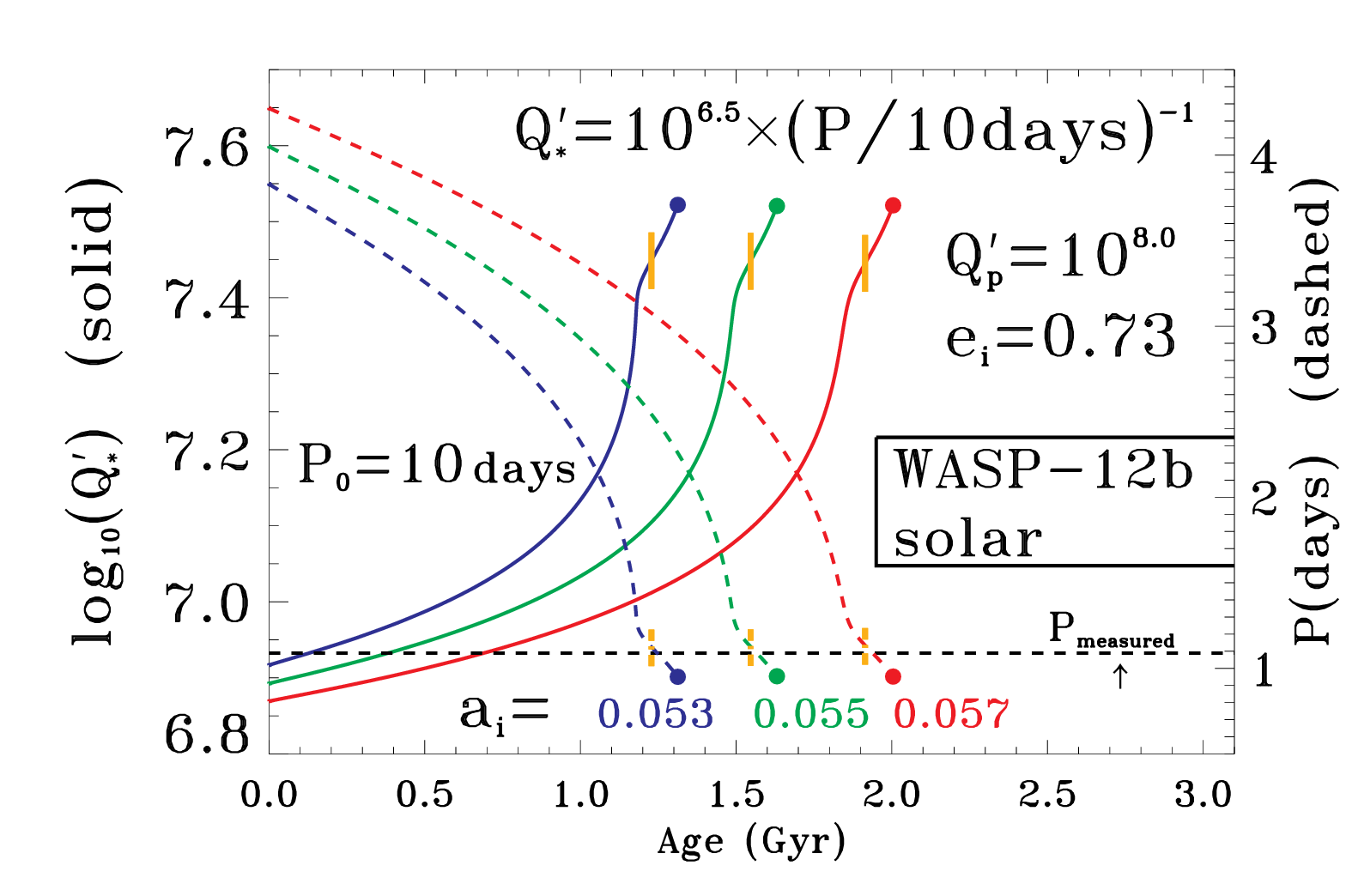}}
%%\rotatebox{90}{test}
\caption{\footnotesize  Same as Fig.\ref{fig:fig1}, but for WASP-12b.
         Note the very narrow range in age ($\sim$50 Myrs) for which the fits are simultaneously obtained for $R_{p}$, $e$, and $a$.
         Note also how close the orbit of the planet is to the Roche limit, which is roughly 0.021~AU. Given the uncertainties in the determination of the semimajor axis $a$ and
	 the orbital eccentricity, the periastron $p=a(1-e)$ of the orbit might be slightly smaller than the Roche limit. We find $0.0207 \lesssim p \lesssim 0.0229$.
         See Section \ref{sec:applications} for more discussion.
         }
\label{fig:fig2}
\end{figure}

\clearpage
\end{landscape}

% |--- fig 3 ---|
\clearpage
%\begin{landscape}
\begin{figure}[ht]
%\plotone{ms_evolution_fig1.eps}
%\centering\includegraphics[scale=0.80]{old_eps/ms_evolution_fig1.eps}
\centerline{
\includegraphics[width=17.0cm,angle=0,clip=true]{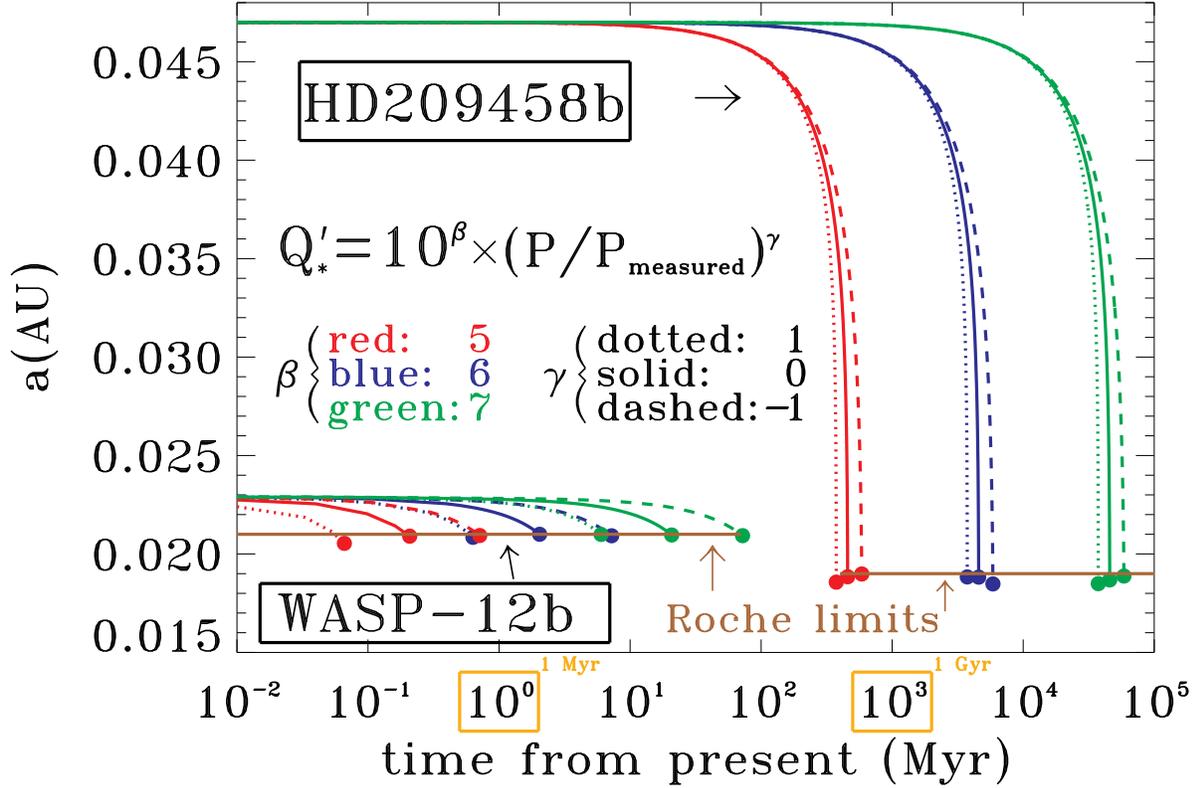}}
\caption{\footnotesize  Semimajor axis $a$(AU) evolution of transiting EGPs driven by tides in the star.
         We assume tidal evolution of $a$ at constant planet and star properties.
         The emphasis is on the dependence of this evolution on the tidal dissipation factor in the star 
         $Q'_{\ast}$, which is assumed to vary with the orbital period $P$, as 
         $Q'_{\ast}=10^{\beta} \times \left(P/P_{\rm measured}\right)^{\gamma}$,
         where $P_{\rm measured}$ is the measured orbital period, $\beta$ and $\gamma$ are \textit{a priori} unknown factors 
         quantifying the evolution of $Q'_{\ast}$.
         We consider three possibilities for $\beta$: 5(red), 6(blue), 7(green), and three possibilities for $\gamma$: 1(dotted), 0(solid), -1(dashed).
         These choices enable us roughly to encompass current observational and theoretical estimates of $Q'_{\ast}$.
         The effect of the tidal dissipation factor in the planet $Q'_{p}$ is negligible in these cases because of the low orbital eccentricities.
         We choose HD~209458b and WASP-12b, two planets evolving on disparate timescales. 
         The planets end up falling into their parent stars and get disrupted at their respective Roche limits (indicated on the figure).
         However, the curves show the very large uncertainties in the timescale to reach this end.
         WASP-12b can plunge in between 0.1 and 100 Myr from now, a 3-order-of-magnitude range. HD~209458b can plunge in between 0.5 and 60 Gyr from now, a 2-order-of-magnitude range.
         The main source of uncertainty comes from the $\beta$ factor.
	 To summarize, this figure demonstrates how difficult it is to predict the evolution of the orbits of transiting EGPs, given the very poor knowledge of the tidal dissipation
         factors in the host stars.
	 See Section \ref{sec:applications} for more discussion.
         }
\label{fig:fig3}
\end{figure}

\clearpage

% ------------------------------------------------------------------------------------------------------------------------------------------------
% ---------------------------------------------------------- end figures -------------------------------------------------------------------------
% ------------------------------------------------------------------------------------------------------------------------------------------------

\end{document}